\documentclass[journal=jctcce,manuscript=article]{achemso}
  
\usepackage[utf8]{inputenc}
\usepackage{amsfonts}
\usepackage{amsbsy}
\usepackage{amssymb}
\usepackage{amsmath}
\usepackage{mathtools}
\usepackage{textcomp}
\usepackage{mathabx}
\usepackage{bm}

\usepackage[title]{appendix}

\newcommand{\brr}[1]{\left( #1 \right) }
\newcommand{\brs}[1]{\left[ #1 \right] }
\newcommand{\brc}[1]{\left\{ #1 \right\} }
\newcommand{\brp}[1]{\left| #1 \right| }
\newcommand{\bapa}[2]{\big< #1 \big| #2 \big> }
\newcommand{\bappa}[3]{\big< #1 \big| #2 \big| #3 \big> }
\newcommand{\bs}[1]{\boldsymbol{#1}}
\newcommand{\mr}[1]{\mathrm{#1}}
\newcommand{\mb}[1]{\mathbf{#1}}
\newcommand{\mc}[1]{\mathcal{#1}}

\author{Stanislav Komorovsky}
\affiliation[SAV]{Institute of Inorganic Chemistry, Slovak Academy of Sciences, D\'ubravsk\'a cesta 9, SK-84536 Bratislava, Slovakia}
\alsoaffiliation[UNI]{Department of Inorganic Chemistry, Faculty of Natural Sciences, Comenius University, Ilkovicova 6, SK-84215 Bratislava, Slovakia}
\email{stanislav.komorovsky@savba.sk}

\title{The existence and unambiguity of the principal axis system of the EPR tensors}

\begin{document}

%
%
\begin{abstract}
Although the role of the electron paramagnetic resonance (EPR) g-tensor and
hyperfine coupling tensor in the EPR effective spin Hamiltonian is discussed
extensively in many textbooks, certain aspects of the theory are missing.  In
this text we will cover those gaps and thus provide a comprehensive theory
about the existence of principal axes of the EPR tensors. However, an important
observation is that both g- and a-tensors have two sets of principal axes---one
in the real and one in the fictitious spin space---and, in fact, are not
tensors.  Moreover, we present arguments based on the group theory why only
eigenvalues of the G-tensor, $\mb{G} = \mb{g}\mb{g}^{\!\mathsf{T}}$, and the
sign of the determinant of the g-tensor are observable quantities (an
analogical situation also holds for the hyperfine coupling tensor). We keep the
number of assumptions to a minimum and thus the theory is applicable in the
framework of the Dirac--Coulomb--Breit Hamiltonian and for any spatial symmetry
of the system.
\end{abstract}

%
%
\section{Notation}
\label{sec:notation}

In this work we employ the Hartree system of atomic units, so for example the
Bohr magneton has the value $\tfrac{1}{2c}$, where $c$ represents the speed of
light.  Furthermore, we assume summation over repeated indices; however, this
rule may be broken in certain instances: for example, when writing the
eigenvalue equation $\mb{A}\mb{C}_k=e_k\mb{C}_k$, then $k$ is not a summation
index. Such cases can be recognized by the same index appearing on both side of
the equation. Nevertheless, in cases where the violation of the summation rule
is not obvious, we state explicitly which indices are not summed over.  Also in
this work the symbols $\mc{R}$ and $\mc{I}$ represent the real and imaginary
part of a complex number, respectively. Finally, variables in bold font
represent either matrices or vectors whose dimension is apparent from the text.

%
%
\section{Introduction}
\label{sec:introduction}

The main aim of this work is to find the principal axis of the g-tensor with no
loose ends, {\it i.e.} all assumptions must be stated and all statements proven.
We will not specify the level of theory for the Hamiltonians and wavefunctions,
in order to make the discussion as general as possible. We make only these two
assumptions:
\begin{enumerate}

  \item
  The Hamiltonian is time-reversal symmetric in the absence of magnetic fields,
  and the Hamiltonian responsible for the interaction with magnetic field is
  time-reversal antisymmetric.

  \item
  The quantum-mechanical (qm) Hamiltonian, expressed in a subspace of
  orthonormal wavefunctions (populated states), can be fully mapped to an
  effective spin Hamiltonian of the form
  \begin{equation}
    \label{eq:eff:gt}
    \mb{H}^\mr{qm} \overset{!}{=} \mb{H}^\mr{eff} = \frac{1}{2c} B_u g_{uv} \mb{S}_v.
  \end{equation}

\end{enumerate}
The first of these assumptions holds, and makes the following discussion valid,
for the Dirac--Coulomb--Breit (DCB) Hamiltonian and any approximate theory that
can be derived from it. The discussion is thus appropriate for the vast
majority of {\it ab initio} quantum chemical methods in use at the present
time. The second assumption is warranted only for systems with negligible
zero-field splitting (ZFS), or for doublet systems where ZFS vanishes. In
addition, under assumption 1, the effective spin Hamiltonian in
eq~\eqref{eq:eff:gt} completely describes the field-dependent part of
$\mb{H}^\mr{qm}$ only up to triplet multiplicity. For systems with higher
multiplicities, eq~\eqref{eq:eff:gt} is valid only if higher than linear
spin--orbit effects are negligible (the so-called weak spin-orbit limit).  It
is therefore only for systems with doublet multiplicity that point 2 holds for
{\it any} strength and order of relativistic effects. As a result of this, and
because every element of the SU$(2)$ group can be expressed using the
exponential parametrisation presented in eq~\eqref{eq:USn} with
$\vec{\mb{S}}=\tfrac{1}{2}\vec{\bs{\sigma}}$, the doublet case is fully
described by the theory discussed in this work. For a more detailed discussion
on the validity of eq~\eqref{eq:eff:gt} see ref~\citenum{Griffith:1960:spinH}.
Finally, note that because the operator associated with the magnetic moment of
the nucleus is time-reversal antisymmetric, the following discussion is also
valid for the hyperfine coupling tensor.

In the rest of this section we will focus on the theory behind the effective
spin Hamiltonian. In the general case, the system in the presence of the
magnetic field is fully described by the many-electron Hamiltonian and its
eigenspectrum ($k$ is not a summation index)
\begin{gather} \label{HPsiEPsi}
  \left( H^0 + H^Z \right) \Psi_k = \mc{E}_k \Psi_k.
\end{gather}
Here $H^0$ is the perturbation-free Hamiltonian and $H^Z$ represents the
Hamiltonian responsible for the interaction of the system with the magnetic
field $\vec B$.  In the following discussion we will assume that the solution
of the eigenproblem for the system in the absence of an external perturbation
is known ($k$ is not a summation index)
\begin{gather} \label{eigen:H0}
  H^0 \Phi_k = E_k \Phi_k.
\end{gather}
The eigenfunctions $\Phi_k$ of the Hamiltonian $H^0$ form an orthonormal
complete basis in the corresponding Hilbert space, so it is convenient to
express the unknown wavefunctions $\Psi_k$ in the basis of those
eigenfunctions.  Finding the approximate solution of the eigenproblem
\eqref{HPsiEPsi} then amounts to constructing the perturbation operator $H^Z$
in the truncated basis of perturbation-free wavefunctions $\Phi_k$ and solving
the following eigenproblem (see Appendix \ref{app:diag})
\begin{gather} \label{eigen:mat}
  \left( \mb{E} + \mb{H}^Z \right) \mb{C} = \mb{C} \bs{\mc{E}},
  \\
  \label{eigen:mat2}
  \left(H^Z\right)_{mn} = \big< \Phi_m \big| H^Z \big| \Phi_n \big>,
  \qquad
  \Psi_k = C_{nk} \Phi_n,
  \\
  \label{eigen:mat3}
  E_{mn} = E_m \delta_{mn},
  \qquad
  \mc{E}_{mn} = \mc{E}_m \delta_{mn},
\end{gather}
where the summation convention is not applied in eq~\eqref{eigen:mat3}.

The basic idea of the effective spin Hamiltonian is to properly describe the
lowest energy levels of the system under study, $\mc{E}_m$, in
eq~\eqref{eigen:mat}.  For this purpose one must find a suitable basis,
$\Phi_n$ in eq~\eqref{eigen:mat2}, onto which to project the many-electron qm
Hamiltonian $H^\mr{qm}$.  Note that the basis may be formed by the
eigenfunctions $\Phi_n$, or any unitary linear combinations of them---in which
case the matrix $\mb{E}$ in eq~\eqref{eigen:mat} is not diagonal.  The target
energy levels are then calculated as eigenvalues of the resulting matrix,
$\mb{H}^\mr{qm}=\mb{E} + \mb{H}^Z$.  The role of the effective spin Hamiltonian
is then to parametrize that matrix:
\begin{gather}
  \label{qm-to-eff}
  \mb{H}^\mr{qm} \overset{!}{=} \mb{H}^\mr{eff},
\end{gather}
wherein the g-tensor---and possibly D-tensor and A-tensor---plays the role of
the expansion coefficients, and the matrices $\mb{S}_v$ that of the basis
elements. The energy levels that one needs to properly describe an EPR
experiment are usually determined by the population---{\it i.e.} the states
that are thermally populated in eq~\eqref{HPsiEPsi}---while the basis,
$\Phi_n$, in which the qm Hamiltonian may be expressed is determined by the
energy differences between the populated energy levels and the energies
$E_n$---thus one may discard basis functions $\Phi_n$ that have large energy
differences. In cases where the nonrelativistic ground-state multiplet is well
separated from the excited states---the usual case for transition metal
complexes---the number of basis functions and of populated states is the same.
On the other hand, if the energy spectrum of the studied systems does not have
a large gap between populated states and other excited states---which is often
the case in systems containing Lanthanides or Actinides---the number of basis
functions should be larger than the number of populated states.

Finally note that some of the theory presented in this work can be found in
refs~\citenum{Messiah-book, Abragam1970, Chibotaru:2013:spinH}.

%
%
\section{Construction of the projected Zeeman Hamiltonian}
\label{sec:gt:construct}

Although the theory discussed in this section is applicable for any
multiplicity of the fictitious spin---under the assumptions given in
section~\ref{sec:introduction}---we will demonstrate various aspects of the
theory using the doublet case as an example.

Equation~\eqref{eq:eff:gt} represents the mapping of the projected Zeeman
Hamiltonian $\mb{H}^\mr{qm}$ to a fictitious (effective) spin space, where the
matrices $\vec{\mb{S}}$ are fictitious spin operators $\vec{S}$ projected on
the fictitious spin basis $\left|S,M\right>$, with $S$ being the fictitious
spin and $M=-S, (-S+1), \dots, S$.  Here we have repeated the word "fictitious"
a little too much, but we want to leave no doubt that when talking about the
spin in this work we refer to the fictitious (effective) spin and not to the
real spin of the system.  In many cases the multiplicity of the effective spin
Hamiltonian, $m=2S+1$, corresponds to the spin multiplicity of the
nonrelativistic ground state of the studied system before it is split by the
spin-orbit interaction.  However, in the general case, $m$ represents the
dimension of $\mb{H}^\mr{qm}$, {\it i.e.} the number of states that are
included in the effective spin Hamiltonian.  One example where the dimension of
the matrices might not correspond to the multiplicity of the nonrelativistic
ground state is a system where the EPR effective spin Hamiltonian describes a
doubly-degenerate ground state and a doubly-degenerate excited state. This
scenario might arise as a zero-field-split nonrelativistic quartet ground state
or it might originate from a doublet ground state with a low-lying doublet
excited state.  Anyhow, both scenarios are described by the fictitious spin
$S=\tfrac{3}{2}$.  To avoid possible confusion, note that whenever we refer to
the multiplicity of a system ({\it e.g.} a doublet case or a triplet system),
we refer to the number of states that are described by the effective spin
Hamiltonian, rather than to the spin multiplicity of the nonrelativistic ground
state.

The integer fictitious spin---$\mb{H}^\mr{eff}$ with odd dimension, describing
an odd number of states---can only describe systems with an even number of
electrons, because systems with odd number of electrons have every energy level
even-times degenerate (see Appendix~\ref{app:odd-electrons}) in which case
$\mb{H}^\mr{eff}$ must have even dimension. On the other hand, half-integer
fictitious spin---$\mb{H}^\mr{eff}$ with even dimension, describing an even
number of states---can in principle be used to describe systems with either an
even or an odd number of electrons. However, a half-integer fictitious spin
describing a system with an even number of electrons is a rather unusual
scenario. Let us consider an example where the nonrelativistic triplet ground
state of a system with axial symmetry is split by strong SO interaction into
two degenerate states with lower energy (a new doubly-degenerate ground state)
and one non-degenerate state with higher energy (which, from the point of view
of qm theory, counts as a non-degenerate excited state). In such a case, if the
ZFS is large enough, it is sufficient for the effective spin Hamiltonian to
describe only the doubly-degenerate ground state.  This ground state is formed
by a so-called non-Kramers pair (see Appendix~\ref{app:non-kramers-partner}),
and it has an EPR spectrum with unusual properties because the eigenfunctions
of a system with an even number of electrons are non-magnetic, see the
discussion in Appendices~\ref{app:non-magnetic1},~\ref{app:non-magnetic2},
and~\ref{app:nonKramers-off} and in ref~\citenum{Griffith:1963:non-KP}.

In this section we will construct the matrix $\mb{H}^\mr{qm}$, used in
eq~\eqref{eq:eff:gt}, as a projection of the Zeeman Hamiltonian onto the space
defined by a set of wavefunctions $\{\Phi_k\}_{k=1}^m$.  In the previous
section the wavefunctions $\Phi_k$ denoted eigenfunctions of $H^0$, but in the
following we will also consider any unitary linear combinations of these
eigenfunctions. In other words, we will consider any other orthonormal set of
wavefunctions that defines the same space. From the point of view of
eq~\eqref{eq:eff:gt} this amounts to a choice of the basis for the projection
of the qm Zeeman operator, and thus it should not change any physical
properties.  However, because $\mb{H}^\mr{eff}$ employs the spin operators
$\vec{\mb{S}}$, which in turn follow specific rules under time-reversal
symmetry, not every choice of basis is suitable for the projection. For
example, in a triplet system of C1 spatial symmetry every eigenfunction of
$H^0$ is non-degenerate and non-magnetic (up to a phase factor, see also the
discussion in Appendix~\ref{app:non-magnetic1}). However, if one wants to use
the expansion in eq~\eqref{eq:eff:gt}, the qm Hamiltonian must be expressed in
a basis where two wavefunctions form a non-Kramers pair and one wavefunction
has the property $\mc{K}\Psi=-\Psi$ under time-reversal symmetry.  Here,
$\mc{K}$ represents the many-electron time-reversal operator, see
Appendix~\ref{app:K}.  Note that, in this example, the wavefunctions are
eigenfunctions of $H^0$ only if the system has axial symmetry.  To choose a
proper basis for the projection of the qm Hamiltonian one should consider the
behavior of the fictitious spin eigenfunctions under time-reversal symmetry.
For fictitious spin $S$ the eigenfunctions of the fictitious spin operators
$S^2$ and $S_z$ follow the relation\cite{Abragam1970}
\begin{gather}
  \label{eq:TRrelations}
  \mc{K}\left|S,M\right> = (-1)^{S-M} \left|S,-M\right>.
\end{gather}
For simplicity we have used the symbol $\mc{K}$ here, although we do not
specify the representation of eq~\eqref{eq:TRrelations}---{\it i.e.} we do not
specify the space in which the spin states $\left|S,M\right>$ are represented.
However, in every other instance in this text we use the symbol $\mc{K}$
strictly to mean the qm many-electron time-reversal operator that acts in the
$S^- H^{\otimes N_\mr{e}}$ Fock subspace for $N_\mr{e}$ fermions (see
Appendix~\ref{app:K}). This minor inconsistency in notation appears only once
in this work, in eq~\eqref{eq:TRrelations}. By imposing eq~\eqref{eq:eff:gt} we
assume a correspondence between the qm and the fictitious spin bases $\Phi_k
\leftrightarrow \left|S,M\right>$, so the qm basis $\{\Phi_k\}_{k=1}^m$ must
follow the relation in eq~\eqref{eq:TRrelations} as well. For a detailed
discussion and expressions on this topic see section III in
ref~\citenum{Chibotaru:2013:spinH}.  One way to construct the basis
$\{\Phi_k\}_{k=1}^m$ that satisfies relation~\eqref{eq:TRrelations} is to
follow this procedure:
\begin{itemize}
  \item
  {\bf For an odd number of electrons:} a) fix the arbitrariness ({\it e.g.}
  phase factors) in the eigenfunctions of $H^0$ such that each energy level is
  represented only by Kramers pairs, see Appendix~\ref{app:odd-electrons}; b) fix
  the phase factors of the selected Kramers partners such that they follow
  eq~\eqref{eq:TRrelations}---note that the choice of which Kramers partner gets
  the phase $-1$ is arbitrary.
  \item
  {\bf For an even number of electrons:} a) make the eigenfunctions of $H^0$
  non-magnetic according to the procedure described in
  Appendix~\ref{app:non-magnetic1}; b) arbitrarily choose one eigenfunction to
  remain non-magnetic up to a phase factor dictated by eq~\eqref{eq:TRrelations};
  c) from the rest of the non-magnetic wavefunctions, construct non-Kramers
  partners as described in Appendix~\ref{app:non-kramers-partner} and fix their
  phases according to eq~\eqref{eq:TRrelations}.
\end{itemize}
Finally, note that using this procedure for a system with an odd number of
electrons results in a basis that consists of eigenfunctions of $H^0$, but for
a system with an even number of electrons this is not the case, because the
procedure described in Appendix~\ref{app:non-kramers-partner}
[eq~\eqref{nonKP}] may mix eigenfunctions that belong to different energy
levels.

As an example we will examine a system with a doubly-degenerate ground state
and well-separated excited states. For such a doublet system the proper qm
basis contains one \mbox{(non-)Kramers pair}
\begin{gather}
  \Phi \coloneqq \Phi_1,
  \\
  \widebar\Phi = \mc{K} \Phi \coloneqq \Phi_2.
  \label{KramersPair2}
\end{gather}
The eigenproblem presented in eq~\eqref{eigen:mat} then reduces to the
following form
\begin{gather} \label{eigen:doublet}
  \left(\begin{array}[c]{cc}
  \big<         \Phi \big| H^Z \big| \Phi \big> & \big<         \Phi \big| H^Z \big| \widebar\Phi \big> \\
  \big< \widebar\Phi \big| H^Z \big| \Phi \big> & \big< \widebar\Phi \big| H^Z \big| \widebar\Phi \big>
  \end{array}\right)
  \left(\begin{array}[c]{c}
  C_{1k}
  \\
  C_{2k}
  \end{array}\right)
  =
  e_k
  \left(\begin{array}[c]{c}
  C_{1k}
  \\
  C_{2k}
  \end{array}\right).
\end{gather}
The eigenvalues $e_k$ represent the magnetically-induced splitting relative to
the ground-state energy in the absence of a magnetic field, $e_k = \mc{E}_k -
E$, $E \coloneqq E_1 = E_2$. In this example, we assume that the wavefunctions
of a non-Kramers pair would be eigenfunctions of $H^0$ with the same energy $E$
({\it e.g.} in a system with an axial symmetry).  Equation
\eqref{eigen:doublet} can be further simplified when considering the
time-reversal and Hermitian symmetry of the Hamiltonian $H^Z$, see
Appendix~\ref{app:Kramers-spin}
\begin{gather} \label{eigen:2:mod}
  \left(\begin{array}[c]{cr}
  \big<         \Phi \big| H^Z \big| \Phi \big> & \big< \Phi \big| H^Z \big| \widebar\Phi \big> \\
  \big< \widebar\Phi \big| H^Z \big| \Phi \big> &-\big< \Phi \big| H^Z \big|         \Phi \big>
  \end{array}\right)
  \left(\begin{array}[c]{c}
  C_{1k}
  \\
  C_{2k}
  \end{array}\right)
  =
  e_k
  \left(\begin{array}[c]{c}
  C_{1k}
  \\
  C_{2k}
  \end{array}\right).
\end{gather}
Also note that the off-diagonal elements are zero in the case of an even number
of electrons---{\it i.e.} for a \mbox{non-Kramers} pair---see
Appendix~\ref{app:nonKramers-off}. As discussed in
section~\ref{sec:introduction} the $2 \times 2$ matrix in eq
\eqref{eigen:2:mod} can be parametrized by the g-tensor as follows
\begin{align}
  \label{gt:def1}
  \frac{1}{4c}
  B_u g_{uv} \sigma_v
  &\overset{!}{=}
  \left(\begin{array}[c]{cr}
  \big<         \Phi \big| H^Z \big| \Phi \big> & \big< \Phi \big| H^Z \big| \widebar\Phi \big> \\
  \big< \widebar\Phi \big| H^Z \big| \Phi \big> &-\big< \Phi \big| H^Z \big|         \Phi \big>
  \end{array}\right),
\end{align}
where $B_u$ are the components of the external magnetic field, and $\sigma_v$
is the Pauli matrix for direction $v$. The individual components of the
g-tensor can then be derived from the elements of the $2 \times 2$ matrix in eq
\eqref{gt:def1}
\begin{align} \label{gt:sol}
  \begin{array}[c]{rr}
  g_{u1} =& 4c\,\mc{R} \bappa{\Phi}{H^Z_u}{\widebar\Phi},
  \\
  g_{u2} =&-4c\,\mc{I} \bappa{\Phi}{H^Z_u}{\widebar\Phi},
  \\
  g_{u3} =& 4c\,\bappa{\Phi}{H^Z_u}{\Phi},
  \end{array}
\end{align}
where $H^Z = B_u H^Z_u$ and
\begin{equation}
  \label{eq:HZu}
  H^Z_u \coloneqq \left.\frac{d H^Z} {d B_u} \right|_{\vec B = 0}.
\end{equation}
An interesting observation is that, on changing the phase of the wavefunction
$\Phi$, the g-tensor elements $g_{u1}$ and $g_{u2}$ change as well, see
Appendix~\ref{app:phase}. This somewhat surprising result leads to the
conclusion that the g-tensor is not an observable quantity in the fullest sense
of the term. Below we will discuss how the observable quantities derived from
the g-tensor are the eigenvalues of $\mb{G} = \mb{g}\mb{g}^\mathsf{T}$ and the
sign of the g-tensor determinant. 

To obtain the g-tensor for higher multiplicities using the same procedure as
outlined for the doublet system, the system must satisfy the assumptions
discussed in section~\ref{sec:introduction}. To obtain the energy splittings
one must in addition assume that all energy levels are degenerate in the
absence of a magnetic field, see eqs~\eqref{eigen:mat}
and~\eqref{eigen:doublet} and the discussion under eq~\eqref{eigen:doublet}. In
other words, in this paragraph we also assume that ZFS effects are negligible
and that there are no other low-lying excited states that must be taken into
account. This is, however, fully warranted only in the doublet case if the
system has an odd number of electrons, where each energy level must be at least
doubly-degenerate even in the relativistic domain, see Appendix~\ref{app:KP}
and the discussion in section~\ref{sec:introduction}.  Then, to find the
splitting of the energy levels induced by a magnetic field one needs to solve
the eigenvalue equation
\begin{equation} \label{eigen:S:mod}
  \left( \frac{1}{2c} B_u g_{uv} \mb{S}_v \right) \mb{C}_k = e_k \mb{C}_k,
\end{equation}
which, for a doublet system, becomes eq~\eqref{eigen:2:mod} with
$\mb{H}^\mr{qm}$ parameterized according to eq~\eqref{gt:def1}. One can now
solve eq~\eqref{eigen:S:mod} (see Appendix~\ref{app:diag-GT}) to get the
expression for the eigenvalues $e_k$ in terms of the g-tensor parameters
\begin{gather} \label{eigen:bS:B}
  e_k = M_k \big| \,\vec b\, \big|,
  \quad
  b_v = \frac{1}{2c} B_u g_{uv},
  \quad
  M_k \in \{-S, -(S-1), \dots, S\},
  \quad
  k = 1,\dots,2S+1.
\end{gather}
In contrast to Appendix~\ref{app:diag-GT} we have here reordered the
eigenvalues in ascending order.  Eq~\eqref{eigen:bS:B} simplifies if the
magnetic field is chosen along the $u$th Cartesian coordinate axis as follows
\begin{align} \label{eigen:bS:u}
  \vec{B}_u = B \vec{n}_u
  \quad \rightarrow \quad
  e_k\big(\vec{B}_u\big) = \frac{M_k B}{2c} \sqrt{ g_{u1}^2 + g_{u2}^2 + g_{u3}^2 },
\end{align}
with $B$ being the size of the magnetic field and $\vec{n}_u$ being the unit
vector in the $u$th direction.  The eigenvalues $e_k$ are observable physical
quantities as they correspond to energy levels in the presence of the magnetic
field $\vec B$, but the form of eq~\eqref{eigen:bS:u} does not allow us to say
the same of the g-tensor parameters. In fact, as was noted above and as we will
see in the next section, the g-tensor is not actually an observable physical
quantity, and moreover is not even a proper tensor.

To conclude the doublet case discussed in this section, the effect of the
magnetic field on the system is fully described by eq~\eqref{eigen:2:mod}
provided only that these two conditions be met: a) $H^0$ is time-reversal
symmetric $H^0$; and b) $H^Z$ is time-reversal antisymmetric. Furthermore, the
g-tensor parameters fully describe the doublet system and are calculated as
presented in eq~\eqref{gt:sol}. Note that this discussion does not depend on
the level of theory used for calculation of the matrix elements $\big< \Phi
\big| H^Z \big| \Phi \big>$ and $\big< \Phi \big| H^Z \big| \widebar\Phi
\big>$, and is therefore valid in, for example, CAS or the DFT class of {\it ab
initio} theories.

%
%
\section{Principal axes of the g-tensor}
\label{sec:principal}

The only physical quantity in the definition of the g-tensor,
eq~\eqref{eq:eff:gt}, is an external magnetic field. The spin matrices,
$\vec{\mb{S}}$, are merely a basis used for the parametrization of the Zeeman
matrix, $\mb{H}^\mr{qm} = \mb{H}^Z$.  Then, the rotation of the laboratory
frame changes only the orientation of the magnetic field, because elements of
$\mb{H}_u^Z$---constructed as the inner product of $H_u^Z$ in some basis, see
{\it e.g.} eq~\eqref{gt:sol}---are invariant with respect to this rotation.
Thus, as we will see later in the discussion, the task of looking for the
principal axes of the g-tensor merely by rotation of the laboratory frame---or,
equivalently, by rotation of the magnetic field---is in the general case
futile.  This is also connected to the interesting observations that the
g-tensor is not a tensor, and moreover that it is not an observable quantity.
The easiest way to understand why the full g-tensor is not an observable
quantity is to see $\vec{\mb{S}}$ as a basis, and the g-tensor as the
associated set of expansion coefficients, with which to represent the Zeeman
matrix $\mb{H}_u^Z$. This matrix is the projection of the Zeeman operator for
the $u$th direction of the magnetic field onto a subspace in a Hilbert space of
$N_\mr{e}$ electrons. Its form of $\mb{H}_u^Z$, however, depends on the choice
of the basis for that subspace. Therefore, the g-tensor itself depends on the
choice of the basis, and a change in basis is obviously not connected to any
observable quantity. In the following, we show that changing this basis rotates
the fictitious spin, and that the g-tensor can be diagonalized when the
appropriate real and fictitious coordinate axis systems are chosen.

Let us first consider the transformation of the g-tensor caused by a change in
the magnetic field. Because the g-tensor does not depend on the strength of the
magnetic field, we consider only different orientations of magnetic fields of
the same strength, $\big|\vec B \big| = \Big| \vec{\widetilde{B}} \Big|$
\begin{equation}
  \label{eq:BtB}
  \vec{\widetilde{B}}
  \quad\rightarrow\quad
  \vec B = \mb{O} \vec{\widetilde{B}}.
\end{equation}
Here $\mb{O}$ is an arbitrary orthogonal transformation, $\mb{O}^\mathsf{T}
\mb{O} = \mb{O} \mb{O}^\mathsf{T} = \mb{1}$, in $\mathbb R^3$, {\it i.e.} $O
\in \mr{O}(3)$. Note that the transformation $\mb{O}$ represents both proper
and improper rotations. Proper rotations, $\mr{det}\,\mb{O} = 1$, have an
exponential form as described in eq~\eqref{rot:R3}, and they form the special
orthogonal group SO$(3)$.  Improper rotations combine a proper rotation and a
reflection, and have $\mr{det} \, \mb{O} = -1$. The g-tensors derived from the
magnetic fields $\vec{\widetilde{B}}$ and $\vec B$ are then connected by the
orthogonal transformation as follows
\begin{gather}
  \mb{H}^\mr{eff}
  =
  \frac{1}{2c} B_u g_{uv} \mb{S}_v
  =
  \frac{1}{2c} O_{uq} \widetilde{B}_q g_{uv} \mb{S}_v
  =
  \frac{1}{2c} \widetilde{B}_u \brr{O_{qu}g_{qv}} \mb{S}_v
  =
  \frac{1}{2c} \widetilde{B}_u \widetilde{g}_{uv} \mb{S}_v,
  \\
  \widetilde{\boldsymbol{g}} = \mb{O}^{\!\mathsf{T}} \!\boldsymbol{g},
  \qquad
  \boldsymbol{g} = \mb{O} \widetilde{\boldsymbol{g}},
  \label{g:trans:B}
\end{gather}
where we have utilized eq~\eqref{eq:BtB} and the fact that $\mb{O}$ is an
orthogonal transformation.

As discussed above, the g-tensor also depends on the choice of the basis onto
which the Zeeman operator is projected. Here we are interested in those linear
transformations of basis functions $\{\Phi_n\}_{n=1}^m$ that preserve their
orthonormality
\begin{gather} \label{U:trans}
  \widetilde\Phi_n = U_{mn} \Phi_m,
  \\
  \bapa{\Phi_m}{\Phi_n} = \delta_{mn}
  \quad
  \Rightarrow
  \quad
  \bapa{\widetilde\Phi_m}{\widetilde\Phi_n} = \delta_{mn}.
\end{gather}
Such matrices are called unitary transformations, $\mb{U}^\dagger \mb{U} =
\mb{U} \mb{U}^\dagger = \mb{1}$, and they act in $\mathbb{C}^m$, {\it i.e.} $U
\in \mr{U}(m)$. Using eq~\eqref{U:trans} one can write the relation between the
elements of the Zeeman Hamiltonian expressed in either of these bases as
\begin{gather}
  \brr{H^Z}_{mn} = \bappa{\Phi_m}{H^Z}{\Phi_n},
  \qquad
  \brr{\widetilde{H}^Z}_{mn} = \bappa{\widetilde\Phi_m}{H^Z}{\widetilde\Phi_n},
  \\
  \bappa{\widetilde\Phi_m}{H^Z}{\widetilde\Phi_n}
  =
  \bappa{U_{km}\Phi_k}{H^Z}{U_{ln}\Phi_l}
  =
  U^\ast_{km} \bappa{\Phi_k}{H^Z}{\Phi_l} U_{ln},
\end{gather}
which can be written in matrix form as
\begin{gather}
  \label{eq:HbarVHV}
  \widetilde{\mb{H}}^Z = \mb{U}^\dagger \mb{H}^Z \mb{U}.
\end{gather}
Each of the Zeeman matrices can be parametrized by a different g-tensor as
follows
\begin{align}
  \label{eq:tildeHZ}
  \widetilde{\mb{H}}^Z
  &\overset{!}{=}
  \frac{1}{2c} B_u \widetilde{g}_{uv} \mb{S}_v,
  \\
  \label{eq:HZ}
  \mb{H}^Z
  &\overset{!}{=}
  \frac{1}{2c} B_u g_{uv} \mb{S}_v.
\end{align}
Equation~\eqref{eq:HbarVHV} allows us to easily transform the effective spin
Hamiltonian between different choices of basis functions that represent the
studied multiplet. Then, by applying eqs~\eqref{eq:tildeHZ} and~\eqref{eq:HZ}
to eq~\eqref{eq:HbarVHV}, one gets
\begin{align}
  \label{eq:tildeg:USU}
  \frac{1}{2c} B_u \widetilde{g}_{uv} \mb{S}_v
  &=
  \frac{1}{2c} B_u g_{uv} \mb{U}^\dagger \mb{S}_v \mb{U}.
\end{align}
Because for every $U \in U(m)$ with $\mr{det}\,\mb{U} = \mr{exp}(i\alpha)$ and
$\alpha \in \mathbb{R}$ there exists $U^+ \in SU(m)$ with $\mr{det}\,\mb{U}^+
= 1$ such that
\begin{gather}
  \label{eq:UpIm}
  \mb{U} = \mb{U}^+ \mb{I}^-_\mr{e},
  \\
  \left(I^-_\mr{e}\right)_{kl} = \delta_{kl} \,e^{i\frac{\alpha}{m}},
  \\
  \mr{det}\,\mb{U}^+ 
  = \mr{det}\brs{ \mb{U} \brr{\mb{I}^-_\mr{e}}^{-1} }
  = \mr{det}\,\mb{U} \,\,\mr{det}\brr{e^{-i\frac{\alpha}{m}} \mb{1} }
  = 1.
\end{gather}
One may transform the right-hand-side of eq~\eqref{eq:tildeg:USU} as
\begin{gather}
  \label{eq:tildeg:UpSUp}
  \frac{1}{2c} B_u \widetilde{g}_{uv} \mb{S}_v
  =
  \frac{1}{2c} B_u g_{uv} \brr{\mb{U}^+}^\dagger \mb{S}_v \mb{U}^+.
\end{gather}
This is possible because $\mb{I}^-_\mr{e}$ is the identity matrix multiplied by
a phase factor.  For $m=2$ the procedure in
eqs~\eqref{U:trans}--\eqref{eq:tildeg:UpSUp} covers all possible $\mb{U}$ and
$\mb{U}^+$, and thus for the doublet case it covers all possible choices of the
orthonormal basis set $\{\Phi_n\}_{n=1}^2$.  For $m>2$ this is not the case,
and therefore in the following we choose only such unitary transformations
$\mb{U}$ in eq~\eqref{U:trans} that lead to $\mb{U}^+$---according to
eq~\eqref{eq:UpIm}---which in turn can be written using the exponential form,
$\mb{U}^+ = \mr{exp}(-i\theta \vec{\mb{S}}\cdot \vec n)$.  For more information
on the topic see also the discussion in Appendix~\ref{app:proper}.  Finally,
one can employ identity~\eqref{SU:SO}---derived in Appendix~\ref{app:rot}---in
eq~\eqref{eq:tildeg:UpSUp}
\begin{gather}
  \label{eq:tildeg:OpS}
  \frac{1}{2c} B_u \widetilde{g}_{uv} \mb{S}_v
  =
  \frac{1}{2c} B_u g_{uv} \left( e^{-i\theta \vec{\mb{R}} \cdot \vec n} \right)_{vw} \mb{S}_w,
  \\
  \label{eq:tildeg:OpS1}
  \bs{\widetilde{g}} = \bs{g} \mb{O}^+,
  \quad
  \mb{O}^+ = e^{-i\theta \vec{\mb{R}} \cdot \vec n}.
\end{gather}
Here the matrix $\mb{O}^+$ is a proper rotation, $\mr{det}\,\mb{O}^+ = 1$, that
is defined by an angle $\theta$ and a unit vector $\vec n$.  The rotation
matrices $\vec R$ are defined in eqs~\eqref{R:def1} and~\eqref{R:def2}, and a
practical prescription for the proper rotation $\mb{O}^+$ is shown in
eq~\eqref{rot:R3}, also known as Rodrigues' formula.  As discussed in
Appendix~\ref{app:eff:spin}, the transformation in eq~\eqref{U:trans} and the
matrix $\mb{O}^+$ in eq~\eqref{eq:tildeg:OpS1} correspond to rotation of the
spin in a fictitious spin space---below we use the notation $\mb{O}^+_\mr{f}$.
On the other hand, the transformation of the magnetic field in
eq~\eqref{eq:BtB} corresponds to rotation in a real space, and thus we denote
it $\mb{O}^+_\mr{r}$.  Therefore, the transformations of the g-tensor,
eqs~\eqref{g:trans:B} and~\eqref{eq:tildeg:OpS1}, are independent and can be
combined as follows
\begin{equation}
  \label{eq:trans:gt}
  \bs{\widetilde{g}} = \mb{O}^{\!\mathsf{T}}_\mr{r} \,\bs{g}\, \mb{O}^+_\mr{f}.
\end{equation}
Note that because these transformations are independent, the choice of which
matrix is or is not transposed is arbitrary. The fact that the transformations
in eq~\eqref{eq:trans:gt} are independent allows us to diagonalize the
g-tensor.  In the following, we show that, under the assumptions presented in
section~\ref{sec:introduction}, the g-tensor can be always brought to a
diagonal form.

Using only the transformation in real space $\mb{O}_\mr{r}$, it is not
possible, in the general case, to diagonalize the g-tensor. It is possible,
however, to diagonalize the $\mb{G}$ matrix instead
\begin{gather}
  \label{G:trans1}
  \mb{G} = \bs{g} \bs{g}^\mathsf{T},
  \\
  \label{G:trans2}
  \mb{G}
  =
  \mb{O}_\mr{r} \, \widetilde{\bs{g}}
  \left( \mb{O}_\mr{r} \, \widetilde{\bs{g}} \right)^{\!\mathsf{T}}
  =
  \mb{O}_\mr{r} \, \widetilde{\bs{g}}
  \, \widetilde{\bs{g}}^\mathsf{T} \mb{O}^{\!\mathsf{T}}_\mr{r}
  =
  \mb{O}_\mr{r} \widetilde{\mb{G}} \mb{O}^{\!\mathsf{T}}_\mr{r}.
\end{gather}
Because $\widetilde{\mb{G}}$ is a real symmetric matrix
\begin{equation}
  \widetilde{\mb{G}}^\mathsf{T}
  =
  (\widetilde{\bs{g}} \widetilde{\bs{g}}^\mathsf{T})^\mathsf{T}
  =
  \widetilde{\bs{g}} \widetilde{\bs{g}}^\mathsf{T}
  =
  \widetilde{\mb{G}},
\end{equation}
there exists an orthogonal transformation $\mb{O}_\mr{r}$ that diagonalizes it,
with the eigenvalues being real numbers. Let us assume that $\mb{O}_\mr{r}$ in
eq~\eqref{G:trans2} is such a transformation---it diagonalizes the
$\widetilde{\mb{G}}$ matrix, and thus $\mb{G}$ has a diagonal form.
$\mb{O}_\mr{r}$ can then be determined by solving the eigenvalue problem.
Following eqs~\eqref{G:trans1} and~\eqref{G:trans2} we can write ($v$ is not a
summation index)
\begin{gather}
  \label{G:diag1}
  \widetilde{G}_{uq} (O_\mr{r})^\mathsf{T}_{qv} = G_{v} (O_\mr{r})^\mathsf{T}_{uv},
  \\
  \label{G:diag2}
  \mb{G}^\mr{diag} = \bs{g} \bs{g}^\mathsf{T},
  \\
  \label{G:diag3}
  \delta_{uv} G_{v} = g_{uq} g_{vq},
  \quad
  G_v \coloneqq G_{vv}.
\end{gather}
Here eq~\eqref{G:diag1} represents the eigenvalue equation with eigenvectors
$\mb{O}^\mathsf{T}_\mr{r}$. The eigenvalues $G_v$ are positive real numbers,
because in eq~\eqref{G:diag3} they are expressed as a sum of the squares of
real numbers ($u$ is not a summation index)
\begin{gather}
  \label{G:diag4}
  G_{u} = g_{uq} g_{uq}.
\end{gather}
Note that $\mb{O}_\mr{r}$ in eq~\eqref{G:diag1} belongs to the group O$(3)$
represented in $\mathbb{R}^3$. However, if $\mr{det}\,\mb{O}_\mr{r} = -1$ one
can simply change the phase of one of the eigenvectors---multiplying the vector
by $-1$---to get the new transformation $\mb{O}^+_\mr{r}$ with $\mr{det}\,
\mb{O}^+_\mr{r} = 1$.  Because changing this phase factor has no physical
consequences, {\it i.e.} one can still diagonalize any $\mb{G}$ matrix, we can
restrict the real transformation in eq~\eqref{eq:trans:gt} to proper rotations
\begin{equation}
  \label{eq:trans:gt1}
  \bs{\widetilde{g}} = (\mb{O}^+_\mr{r})^{\!\mathsf{T}} \,\bs{g}\, \mb{O}^+_\mr{f}.
\end{equation}

Equation \eqref{G:diag3} has two important consequences. First, the eigenvalues
$e_k$ in eq~\eqref{eigen:bS:u} can be expressed using the eigenvalues $G_u$
thanks to their dependence on the g-tensor parameters presented in eq
\eqref{G:diag4}
\begin{gather}
  \label{eigen:G}
  \vec{B}_u = B \vec{n}_u
  \quad \rightarrow \quad
  e_k\big(\vec{B}_u\big)
  = \frac{M_k B}{2c} \sqrt{ g_{u1}^2 + g_{u2}^2 + g_{u3}^2 }
  = \frac{M_k B}{2c} \sqrt{G_u}.
\end{gather}
Now, it is clear that the $\mb{G}$ matrix is an observable physical quantity
and that its eigenvalues govern the splitting induced by the magnetic field
applied along the principal axes of the $\mb{G}$ matrix [see also the
assumptions that led to eq~\eqref{eigen:bS:u} discussed in
section~\ref{sec:gt:construct}].  Moreover, the $\mb{G}$ matrix is a proper
tensor, because under the transformation of the magnetic field (or,
equivalently, the coordinate system) both of its indices are transformed by the
same orthogonal transformation, see eq~\eqref{G:trans2}.  Thus, in the
following, we refer to the $\mb{G}$ matrix as the G-tensor.  This is in
contrast to the g-tensor where only one index is transformed when rotating the
magnetic field, see eq~\eqref{g:trans:B}.  However, one could force the
transformations in eq~\eqref{eq:trans:gt1} to be the same, {\it i.e.} when
rotating the real coordinate system by $\mb{O}^+_\mr{r}$ one could
simultaneously change the basis set $\{\Phi_n\}_{n=1}^m$ such that the
$\mb{O}^+_\mr{f} = \mb{O}^+_\mr{r}$. This procedure would make the g-tensor a
proper tensor.  However, because the g-tensor is in the general case a
non-symmetric matrix, this choice would prohibit its diagonalization.

The second consequence of eq \eqref{G:diag3} is more subtle. The expression
states that if the G-tensor is diagonal, then the rows of the corresponding
g-tensor form an orthogonal set of vectors. This interesting fact can be
exploited in the diagonalization of the g-tensor itself. Let us normalize these
vectors---the rows of the g-tensor---to obtain the orthonormalized set of
vectors $\{\vec{w}_v\}_{v=1}^3$
\begin{align}
  \left(w_v\right)_q = g_{vq} \, G^{-\frac{1}{2}}_v.
  \label{eq:def:W1}
\end{align}
For simplicity let us assume for a while that all $G_v$ values are nonzero,
{\it i.e.} there are no zero rows of the g-tensor, see eq~\eqref{G:diag4}.
Every set of real orthonormal vectors forms an orthogonal transformation [an
element of the group $\mr{O}(m)$], if the number of the vectors equals the
dimension of the vector space---in this case $m$. This statement holds for any
finite-dimensional vector space with a scalar (inner) product. We work in
$\mathbb{R}^3$, but the proof can be easily extended to any complex (or
quaternion) vector space with a finite dimension, see Appendix \ref{app:U} for
more details. We can, therefore, write the following expressions for the set of
vectors $\{\vec{w}_v\}_{v=1}^3$ 
\begin{align}
  \label{eq:def:W2}
  W_{qv} \coloneqq \left(w_v\right)_q
  \quad
  \Rightarrow
  \quad
  \mb{W}^\mathsf{T} \mb{W} = \mb{W} \mb{W}^\mathsf{T} = \mb{1},
  \quad
  \mb{W} \in \mr{O}(3).
\end{align}
Now let us use the orthogonal transformation $\mb{W}$ in eq \eqref{G:diag2} as
follows
\begin{align}
  \label{eq:gWWg}
  \mr{diag}(\mb{G}) = \boldsymbol{g} \boldsymbol{g}^\mathsf{T}
  =
  \boldsymbol{g} \mb{W} \,\mb{W}^\mathsf{T} \!\boldsymbol{g}^\mathsf{T}
  =
  \widebar{\boldsymbol{g}} \widebar{\boldsymbol{g}}^\mathsf{T},
  \qquad
  \widebar{\boldsymbol{g}} = \boldsymbol{g} \mb{W}.
\end{align}
Because of the specific form of the transformation matrix $\mb{W}$,
eqs~\eqref{eq:def:W1} and~\eqref{eq:def:W2}, the $\widebar{g}$-tensor has a
diagonal form with positive diagonal values ($v$ is not a summation index)
\begin{align} \label{g:diag}
  \widebar{g}_{uv} = g_{uq} W_{qv} = g_{uq} \left(w_v\right)_q
  =
  g_{uq} g_{vq} \, G^{-\frac{1}{2}}_v
  =
  \delta_{uv} G^{\frac{1}{2}}_v,
\end{align}
where we have utilized eq~\eqref{G:diag3}. Equations
\eqref{G:diag1}--\eqref{G:diag3} and \eqref{eq:def:W1}--\eqref{g:diag}
represent a recipe for diagonalizing any real non-symmetric square matrix with
two different orthogonal transformations where the final diagonal elements are
positive real numbers. The only assumption was that eigenvalues of the $\mb{G}$
matrix are nonzero. In the case of $G_v$ being zero the corresponding rows of
the g-tensor must be zero as well, see eq~\eqref{G:diag4}. However, the zero
vectors cannot be normalized, so in such a case the set $\{\vec{w}_v\}_{v=1}^3$
does not consist of orthonormal vectors.  A way around this problem is to
replace the zero vectors with ones that will make the set
$\{\vec{w}_v\}_{v=1}^3$ orthonormal. This can be always done, although the
choice of such vectors is not, in the general case, unique. The $\mb{W}$ matrix
is then orthogonal, and when evaluating eq~\eqref{g:diag} one obtains the
expected zero elements---the square root of zero is zero---on the diagonal,
thanks to the corresponding zero rows of the g-tensor. Thus the procedure of
diagonalizing the $\widebar{\bs{g}}$ matrix in eq~\eqref{g:diag} can be
achieved even when some of the $G_v$ values are zero.

To transform the g-tensor, as requested by eq~\eqref{g:diag}, one can use a
rotation in the fictitious spin space, $\mb{O}^+_\mr{f}$, see
eq~\eqref{eq:trans:gt1} and the corresponding discussion.  However, the
determinant of the transformation $\mb{W}$ can be either positive or negative,
while the transformation of the wavefunction basis $\{\Phi_n\}_{n=1}^m$ leads
only to proper rotations, {\it i.e.} $\mb{O}^+_\mr{f}$ with determinant equal
to one. In other words, not all required transformations $\mb{W}$ can be
obtained by mixing the basis set $\{\Phi_n\}_{n=1}^m$. In the problematic case
when $\mr{det} \,\mb{W} = -1$ one can define a new transformation of the
g-tensor as follows
\begin{gather}
  \label{eq:gWI}
  \widebar{\mb{g}} = \mb{g} \mb{W} \mb{I}^-,
\end{gather}
with the matrix $\mb{I}^-$ being any of these matrices
\begin{gather}
  \label{eq:Ichoice}
  \left(\begin{array}[c]{rrr}
  -1 &  0 &  0 \\
   0 &  1 &  0 \\
   0 &  0 &  1
  \end{array}\right),
  \quad
  \left(\begin{array}[c]{rrr}
   1 &  0 &  0 \\
   0 & -1 &  0 \\
   0 &  0 &  1
  \end{array}\right),
  \quad
  \left(\begin{array}[c]{rrr}
   1 &  0 &  0 \\
   0 &  1 &  0 \\
   0 &  0 & -1
  \end{array}\right),
  \quad
  \left(\begin{array}[c]{rrr}
  -1 &  0 &  0 \\
   0 & -1 &  0 \\
   0 &  0 & -1
  \end{array}\right).
\end{gather}
The transformation $\mb{W} \mb{I}^-$ is then a proper rotation and it still
diagonalizes the $\widebar{g}$-tensor
\begin{gather}
  \brr{ \mb{W} \mb{I}^- }^\mathsf{T} \brr{ \mb{W} \mb{I}^- }
  =
  (\mb{I}^-)^\mathsf{T} \mb{W}^\mathsf{T} \mb{W} \mb{I}^-
  =
  (\mb{I}^-)^\mathsf{T} \mb{I}^-
  =
  \mb{1},
  \\
  \brr{ \mb{W} \mb{I}^- } \brr{ \mb{W} \mb{I}^- }^\mathsf{T}
  =
  \mb{W} \mb{I}^- (\mb{I}^-)^\mathsf{T} \mb{W}^\mathsf{T}
  =
  \mb{W} \mb{W}^\mathsf{T}
  =
  \mb{1},
  \\
  \mr{det}\brr{ \mb{W} \mb{I}^- }
  =
  \mr{det}\,\mb{W} \,\mr{det}\,\mb{I}^-
  =
  1,
  \\
  \label{eq:gbar:4}
  \widebar{g}_{uv} = g_{uq} W_{qp} (I^-)_{pv} =
  g_{uq} \left(w_p\right)_q (I^-)_{pv}
  =
  g_{uq} g_{pq} \, G^{-\frac{1}{2}}_p (I^-)_{pv}
  =
  G^{\frac{1}{2}}_u (I^-)_{uv}.
\end{gather}
One can easily show that both g-tensors, $\widebar{\bs{g}}$ and $\bs{g}$, have
the same G-tensor and thus lead to the same energy splittings,
eq~\eqref{eigen:G} ($v$ is not a summation index)
\begin{align}
  \widebar{g}_{uq} \, \widebar{g}_{vq} = g_{uq} \, g_{vq} = \delta_{uv} G_v.
\end{align}
Note that the signs of the determinants of the matrices $\mb{W}$ and $\mb{g}$
are the same, because combining eqs~\eqref{eq:def:W1} and~\eqref{eq:def:W2}
results in
\begin{gather}
  \mr{det}\,\mb{W} = \mr{det}\,\mb{W}^\mathsf{T}
  = \mr{det}\brr{ \mb{G}^{-\frac{1}{2}} \mb{g} }
  = \mr{det}\,\mb{G}^{-\frac{1}{2}} \,\,\mr{det}\,\mb{g},
\end{gather}
where the matrix $\mb{G}^{-\frac{1}{2}}$ is diagonal, with elements
$G_u^{-\frac{1}{2}}$, and its determinant is thus always positive.  Therefore,
if the determinant of the g-tensor is negative, then in the process of its
diagonalization one may choose any matrix $\mb{I}^-$ from
eq~\eqref{eq:Ichoice}.  This choice than means that the signs of the diagonal
elements of the $\widebar{g}$-tensor---also known as the principal elements or
g-values---may be chosen in four different ways while keeping the sign of the
determinant (the product of the g-values) unchanged, see eq~\eqref{eq:gbar:4}.
Therefore, the signs of the individual g-values are not observable quantities.
In the doublet case, there is no degree of freedom left in the transformation
of the basis, eq~\eqref{U:trans}, that could lead to transformations other than
$\mb{O}^+_\mr{f}$ (see also Appendix~\ref{app:proper}), and thus the only
observable quantity is the sign of the determinant of the g-tensor and the
eigenvalues of the G-tensor.  In the case of higher multiplicities, the
additional degree of freedom in the change of the basis could in principle
prohibit even the measurement of the sign of the determinant of the
g-tensor---the theory presented here can neither confirm this nor rule it out.
However, it turns out that the sign of the g-tensor may be measured even for
higher multiplicies, see ref~\citenum{Pryce:1959:GT-sign}.  Finally, note that
the g-tensor has two sets of principal axis systems, one in real space and one
in the fictitious spin space.

We may summarize the procedure above for diagonalizing the g-tensor as follows:
\begin{enumerate}
  \item
  Use independent rotations in real space and fictitious spin space, see
  eq~\eqref{eq:trans:gt}.
  \item
  Diagonalize the $\widetilde{G}$-tensor, $\widetilde{\mb{G}} =
  \widetilde{\bs{g}} \widetilde{\bs{g}}^\mathsf{T}$, according to
  eq~\eqref{G:trans2}, using the transformation $\mb{O}_\mr{r}$ from
  eq~\eqref{G:diag1}.
  \item
  Construct the g-tensor such that $\bs{g} = \mb{O}_\mr{r}\, \widetilde{\bs{g}}$.
  \item
  Construct the orthogonal matrix $\mb{W}$ combining eqs~\eqref{eq:def:W1}
  and~\eqref{eq:def:W2} using the g-tensor
  \begin{equation}
    \mb{W} = \bs{g}^\mathsf{T} \mb{G}^{-\frac{1}{2}},
  \end{equation}
  where $\mb{G}^{-\frac{1}{2}}$ is a diagonal matrix with diagonal elements
  $G_v^{-\frac{1}{2}}$ and $G_v$ represent eigenvalues of the
  $\widetilde{G}$-tensor, {\it i.e.} diagonal elements of the matrix $\mb{G}$ in
  eq~\eqref{G:trans2}.
  \item
Construct the proper rotation in the fictitious spin space
\begin{align}
  \label{eq:gmOf}
  &\mr{det}\,\bs{g} > 0
  \quad
  \Rightarrow
  \quad
  \mb{O}^+_\mr{f} = \mb{W},
  \\
  \label{eq:gpOf}
  &\mr{det}\,\bs{g} < 0
  \quad
  \Rightarrow
  \quad
  \mb{O}^+_\mr{f} = \mb{W} \mb{I}^-,
\end{align}
with $\mb{I}^-$ being one of the matrices in eq~\eqref{eq:Ichoice}. The
resulting diagonal $\widebar{g}$-tensor has the form
\begin{equation}
  \widebar{\bs{g}} = \bs{g} \mb{O}^+_\mr{f},
\end{equation}
with the diagonal elements $G_v^{\frac{1}{2}}$, which in the case of
$\mr{det}\,\bs{g} < 0$ must incorporate a minus sign, according to which matrix
$\mb{I}^-$ was chosen. Finally, note that in the case of $\mr{det}\,\bs{g} >
0$, in principle one could modify the g-values in a similar way as for the case
of $\mr{det}\,\bs{g} < 0$, but with matrices that have two values $-1$ on
the diagonal. This is, however, not done in practice, and all g-values are
chosen to be positive.
\end{enumerate}
This procedure represents a proof of existence, and to a certain degree
uniqueness, of principal axes and principal values of the g-tensor. In
practice, however, the principal values are obtained simply by taking the
square roots of the eigenvalues of the G-tensor (step 2) and by modifying their
sign if the determinant of the g-tensor is negative [see eq~\eqref{eq:gbar:4}].
The two sets of principal axes---in the real space and the fictitious spin
space---are obtained in point 2 and point 5. However, when using {\it ab
initio} wavefunction methods one may be interested in the basis
[eqs~\eqref{U:trans}--\eqref{eq:HbarVHV}] in which the g-tensor is diagonal.
This is a more challenging task, because it requires extraction of $\theta$ and
$\vec n$ from eq~\eqref{eq:gmOf} or~\eqref{eq:gpOf} with $\mb{O}^+_\mr{f} =
\mr{exp}(-i\theta \vec{\mb{R}} \cdot \vec n)$ and then calculation of the
unitary transformation in eq~\eqref{U:trans} as $\mb{U} = \mr{exp}(-i\theta
\vec{\mb{S}}\cdot \vec n)$. There is, however, a better way of solving this
problem, which we will discuss in the next section.

%
%
\section{An alternative procedure for the diagonalization of the g-tensor}
\label{sec:alternative}

In this section we will discuss an alternative procedure to the diagonalization
of the g-tensor that will provide us with the basis set
$\{\Phi_n\}_{n=1}^m$---in which the g-tensor is diagonal---without the need for
extracting the parameters $\theta$ and $\vec n$ from the rotation in the
fictitious spin space $\mb{O}^+_\mr{f} = \mr{exp}(-i\theta \vec{\mb{R}} \cdot
\vec n)$, see also the last paragraph in the previous section. This procedure
is commonly used in quantum chemical calculations, see for example
refs~\citenum{Chibotaru:2013:spinH} and~\citenum{Bolvin:2016:EPR-review} and
works cited therein. Although this procedure is more practical then the one
presented in section~\ref{sec:principal}, it does not provide a stand-alone
proof that the g-tensor can be always brought to the diagonal form. On the
contrary, the procedure described here is applicable to an arbitrary system
only if a principal axis system exists for the g-tensor [and even then, only if
the transformation of the g-tensor can be written in the form given in
eq~\eqref{eq:trans:gt}].  However, a proof of the diagonalizability of the
g-tensor was presented in the previous section, and we will use some of the
results from that section here.

In the following we assume that the G-matrix is already diagonal, {\it i.e.} we
have already found the principal axis system in real space. Furthermore, let us
have a starting reference basis set $\{\Phi_n\}_{n=1}^m$ that satisfies the
same transformations under time-reversal symmetry as presented in
eq~\eqref{eq:TRrelations} [see also the corresponding discussion under that
equation].  The Zeeman matrix that corresponds to the $u$th component of the
magnetic field can then be parametrized using the g-tensor as follows
\begin{gather}
  \label{eq:HuZ}
  \mb{H}_u^Z
  \overset{!}{=}
  \frac{1}{2c} g_{uv} \mb{S}_v.
\end{gather}
Applying an arbitrary unitary transformation $\mb{U}$ to the basis set
$\{\Phi_n\}_{n=1}^m$, eq~\eqref{U:trans}, we get
\begin{gather}
  \label{eq:HubarVHV}
  \widetilde{\mb{H}}_u^Z = \mb{U}^\dagger \mb{H}_u^Z \mb{U}.
\end{gather}
This equation corresponds to eq~\eqref{eq:HbarVHV} for individual components of
the magnetic field.  Our goal is to find a unitary transformation of the
reference basis set that diagonalizes the g-tensor in eq~\eqref{eq:HuZ}. In
other words, each of the three matrices, $\widetilde{\mb{H}}_u^Z$, should have
only one nonzero component in their expansion into the basis of spin matrices
$\vec{\mb{S}}$, {\it i.e.} ($u$ is not a summation index)
\begin{gather}
  \label{eq:HuZ:zz}
  \widetilde{\mb{H}}_u^Z
  = \tfrac{1}{2c} \widetilde{g}_{uv} \mb{S}_v
  = \tfrac{1}{2c} \widetilde{g}_{uu} \delta_{uv} \mb{S}_v
  = \tfrac{1}{2c} \widetilde{g}_{uu} \mb{S}_u.
\end{gather}
Because the spin matrix $\mb{S}_3$ is diagonal, one can achieve this goal for
the $z$th Zeeman matrix by diagonalizing it
\begin{gather}
  \label{eq:H3Z}
  \mb{H}_3^Z \mb{C} = \mb{C} \mb{e},
\end{gather}
where the diagonal matrix $\mb{e}$ contains the eigenvalues and the matrix
$\mb{C}$ represents the corresponding eigenvectors.  Because the coefficients
$\mb{C}$ form a unitary transformation, $\mb{C}^\dagger \mb{C} = \mb{C}
\,\mb{C}^\dagger = \mb{1}$ (see Appendix~\ref{app:U}), one can write
\begin{gather}
  \label{eq:H3Z:UC}
  \widetilde{\mb{H}}_3^Z = \mb{C}^\dagger \mb{H}_3^Z \mb{C} = \mb{e},
  \quad
  \mb{U} = \mb{C}.
\end{gather}
The $\widetilde{g}_{33}$ g-value can then be obtained by comparing
eqs~\eqref{eq:HuZ:zz} and~\eqref{eq:H3Z}
\begin{gather}
  \label{eigen:g33:1}
  e_k = \frac{1}{2c} \widetilde{g}_{33} M_k,
  \\
  \label{eigen:g33:2}
  M_k \in \{-S, -(S-1), \dots, S\},
  \quad
  k = 1,\dots,2S+1.
\end{gather}
There are two problems with this result. First, one should determine the
g-value $\widetilde{g}_{33}$ from eq~\eqref{eigen:g33:1}, but the equation must
be satisfied for every index $k$, so the problem is over-parametrized.  The
second problem is that we can only use one unitary transformation for all three
matrices $\widetilde{\mb{H}}_u^Z$. Both problems are a consequence of one more
fundamental issue: is there a unitary transformation that leads to
eq~\eqref{eq:HuZ:zz} for every system that satisfies the assumptions discussed
in section~\ref{sec:introduction}?  The way out of this problem is to consider
section~\ref{sec:principal} as a proof that such a unitary matrix exists. Then,
because the eigenvalues in eq~\eqref{eq:H3Z} are unique, eq~\eqref{eigen:g33:1}
must have the same solution for every $k$.  On the other hand, the eigenvectors
that diagonalize $\mb{H}_3^Z$ are not unique, because when changing the phase
factor of every eigenvector independently, eq~\eqref{eq:H3Z} remains satisfied.
As a result, even though we have carefully chosen the basis set
$\{\Phi_n\}_{n=1}^m$ such that it is possible to parametrize $\mb{H}_u^Z$ as
presented in eq~\eqref{eq:HuZ}, the matrices $\widetilde{\mb{H}}_x^Z$ and
$\widetilde{\mb{H}}_y^Z$ cannot be expressed the same way for every possible
unitary transformation $\mb{U}$, {\it i.e.} with every possible choice of the
phase factors.  One can overcome this problem by just using a unitary
transformation of the form $\mb{U} = \mr{exp}(-i\theta \vec{\mb{S}}\cdot \vec
n)$ , because thanks to the identity in eq~\eqref{SU:SO} the vector
$\vec{\mb{S}}$ is simply rotated when such a transformation is employed [see
also eqs~\eqref{eq:tildeg:UpSUp} and~\eqref{eq:tildeg:OpS}].  Fortunately,
another consequence of the discussion in
sections~\ref{sec:introduction}--\ref{sec:principal} is that the g-tensor can
be diagonalized and that the unitary transformation $\mb{U}$ that is necessary
for the process [see eq~\eqref{U:trans} and related discussion] can be
parametrized using the exponential mapping
\begin{gather}
  \label{eq:exp:U}
  \mb{U} = e^{-i\theta \vec{\mb{S}}\cdot \vec n}.
\end{gather}
In other words, $\mb{U}$ is a member of the $m$-dimensional irreducible
representation of SU$(2)$ [which is a subgroup of SU$(m)$].  This is a helpful
statement, because it means that by proper choice of the phase factors for the
coefficients $\mb{C}$ in eq~\eqref{eq:H3Z}, the unitary transformation $\mb{U}$
in eq~\eqref{eq:H3Z:UC} can be made to satisfy eq~\eqref{eq:exp:U}.  In the
following, we consider the doublet case, $m=2$, separately, because the
exponential form of $\mb{U}$ is easily guaranteed. Further on below we will
discuss the procedure for the general case of arbitrary $m$.

{\bf The doublet case:} As we stated above, the eigenvectors that diagonalize
$\mb{H}_3^Z$ are not unique, because one can change their phase factors freely
and eq~\eqref{eq:H3Z} will remain satisfied
\begin{gather}
  \label{eq:Cpahse1}
  p_{kl} = \delta_{kl} e^{i\alpha_k},
  \quad
  \alpha_k \in \mathbb{R},
  \\
  \label{eq:Cpahse2}
  \mb{p}^\dagger \mb{p} = \mb{p} \,\mb{p}^\dagger = \mb{1},
  \\
  \label{eq:Cpahse3}
  \widetilde{\mb{C}} = \mb{C} \mb{p},
  \quad
  \widetilde{\mb{e}} = \mb{p}^\dagger \mb{e} \mb{p} = \mb{e},
  \\
  \label{eq:Cpahse4}
  \mb{H}_3^Z \widetilde{\mb{C}} = \widetilde{\mb{C}} \mb{e}.
\end{gather}
Because the matrix $\mb{C}$ forms a unitary transformation, its determinant is a
simple phase factor, $\mr{det}\,\mb{C} = \mr{exp}(i\gamma)$, $\gamma \in
\mathbb{R}$ [see eqs~\eqref{phase:U:1}--\eqref{phase:U:4}]. Therefore, one can
always choose the phase factors for the eigenvectors such that the determinant
of the new coefficient matrix $\widetilde{\mb{C}}$ is equal to one
\begin{gather}
  \label{eq:alpah:p}
  \mr{det}\,\mb{C} = e^{i\gamma},
  \quad
  \sum_{k=1}^m \alpha_k = -\gamma
  \quad
  \Rightarrow
  \quad
  \mr{det}\,\widetilde{\mb{C}} = \mr{det}\,\mb{C} \,\mr{det}\,\mb{p} = 1.
\end{gather}
As this procedure can be always performed, without loss of generality we can
assume that the matrix $\mb{C}$ in eq~\eqref{eq:H3Z} already has a determinant
equal to one, and can write
\begin{gather}
  \label{eq:H3Z:SU}
  \mb{U} \in \mr{SU}(2),
  \quad
  \mr{det}\,\mb{U} = 1.
\end{gather}
When representing elements of the group $\mr{SU}(2)$ in the two dimensional
complex space $\mathbb{C}^2$, every group element can be written in the
exponential form as presented in eq~\eqref{eq:exp:U} with $\vec{\mb{S}} =
\tfrac{1}{2}\vec{\sigma}$.  The transformation in eq~\eqref{eq:HubarVHV} then
leads to matrices $\widetilde{\mb{H}}_u^Z$ that can be again parametrized using
only spin matrices with expansion coefficients $\widetilde{\mb{g}}$ [see also
eqs~\eqref{eq:tildeg:OpS} and~\eqref{eq:tildeg:OpS1}]
\begin{gather}
  \label{eq:tildeHgS}
  \widetilde{\mb{H}}_u^Z
  =
  \frac{1}{2c} \widetilde{g}_{uv} \mb{S}_v.
\end{gather}
Therefore one can write
\begin{gather}
  \label{eq:gtogtilde}
  \mb{G}
  = \bs{g} \bs{g}^\mathsf{T}
  = \widetilde{\bs{g}} (\mb{O}^+)^\mathsf{T}
  [ \widetilde{\bs{g}} (\mb{O}^+)^\mathsf{T} ]^\mathsf{T}
  = \widetilde{\bs{g}} \widetilde{\bs{g}}^\mathsf{T}
  = \widetilde{\mb{G}},
\end{gather}
with $\mb{O}^+$ connected to the unitary transformation according to
eq~\eqref{SU:SO:2}.  Furthermore, because we have assumed that the G-tensor is
diagonal, according to eq~\eqref{eq:gtogtilde} the $\widetilde{\mr{G}}$-tensor
is diagonal as well.  Diagonalizing the matrix $\mb{H}_3^Z$ then satisfies
eq~\eqref{eq:HuZ:zz} for $u=3$, so two elements of the $\widetilde{g}$-tensor
are zero
\begin{gather}
  \label{eq:g-zero}
  \widetilde{g}_{31} = \widetilde{g}_{32} = 0.
\end{gather}
In addition, because $\widetilde{\mr{G}}$-tensor is diagonal, the rows of the
$\widetilde{g}$-tensor form an orthogonal set of vectors,
$\{\vec{v}_u\}_{u=1}^3$, see also eq~\eqref{eq:def:W1} and related discussion
\begin{gather}
  (v_u)_v = \widetilde{g}_{uv},
  \\
  \vec{v}_u \cdot \vec{v}_v = 0,
  \quad
  u\ne v.
\end{gather}
Note that $\vec{v}_3$ has the first two components zero, see
eq~\eqref{eq:g-zero}.  Thus, if we further assume that $\widetilde{g}_{33} \ne
0$ then
\begin{gather}
  \label{eq:v3v1}
  \vec{v}_3 \cdot \vec{v}_1 = \widetilde{g}_{33} \,\widetilde{g}_{13} = 0
  \quad
  \Rightarrow
  \quad
  \widetilde{g}_{13} = 0,
  \\
  \label{eq:v3v2}
  \vec{v}_3 \cdot \vec{v}_2 = \widetilde{g}_{33} \,\widetilde{g}_{23} = 0
  \quad
  \Rightarrow
  \quad
  \widetilde{g}_{23} = 0.
  \\
  \label{eq:v1v2}
  \Rightarrow
  \quad
  \vec{v}_1 \cdot \vec{v}_2 = \widetilde{g}_{11} \,\widetilde{g}_{21}
                            + \widetilde{g}_{12} \,\widetilde{g}_{22} = 0.
\end{gather}
Combining eqs~\eqref{eq:g-zero}, \eqref{eq:v3v1} and~\eqref{eq:v3v2} one gets
\begin{gather}
  \label{eq:gtilde:almost}
  \widetilde{\bs{g}} =
  \left(\begin{array}[c]{rrr}
   \widetilde{g}_{11} & \widetilde{g}_{12} &  0 \\
   \widetilde{g}_{21} & \widetilde{g}_{22} &  0 \\
   0 &  0 & \widetilde{g}_{33}
  \end{array}\right).
\end{gather}
As discussed above, diagonalization of $\mb{H}_3^Z$ is (in the general case)
not sufficient to diagonalize the g-tensor. However, one may take advantage of
the remaining degree of freedom in the coefficients in eq~\eqref{eq:H3Z} and
change their phase factors such that the determinant of the matrix $\mb{C}$ is
preserved [see also eqs~\eqref{eq:Cpahse1}--\eqref{eq:Cpahse4}]
\begin{gather}
  \label{eq:pahse:0}
  \mr{det}\,\mb{C} = 1,
  \quad
  p_{kl} = \delta_{kl} e^{i\alpha_k},
  \quad
  \sum_{k=1}^m \alpha_k = 0
  \quad
  \Rightarrow
  \quad
  \mr{det}\,\widetilde{\mb{C}} = \mr{det}\,\mb{C} \,\mr{det}\,\mb{p} = 1.
\end{gather}
In the doublet case, because eq~\eqref{eq:pahse:0} leads only to matrices
$\mb{p}$ that belong to SU$(2)$, the matrix $\widetilde{\mb{C}} = \mb{C}
\mb{p}$ belongs to SU$(2)$ as well and thus can be written in exponential form.
Although this procedure works for the doublet case only, we formulate the
following equations for any multiplicity, to show that one can choose the form
of the matrix $\mb{p}$ such that it belongs to the $m$-dimensional irreducible
representation (irrep) of SU$(2)$. If $\mb{C}$ then belongs to this irrep,
$\widetilde{\mb{C}}$ will belong to the same group as well, {\it i.e.} one can
write it in an exponential form, eq~\eqref{eq:exp:U}. However, as discussed
above, $\mb{C}$ might not belong to this irreducible representation, and thus
one needs to apply a different procedure to diagonalize the g-tensor for higher
than doublet multiplicities.  In the doublet case, eq~\eqref{eq:pahse:0} leads
to the matrix $\mb{p}$ in the following expression without any restrictions,
while for higher multiplicities one must additionally restrict the choice of
the phase factors
\begin{gather}
  \label{eq:pS3}
  \{\alpha_k\}_{k=1}^m = \{ - \eta S, - \eta (S-1), \dots, \eta S \},
  \quad
  m = 2S + 1,
  \quad
  \eta \in \mathbb{R}
  \\
  \label{eq:pS3-1}
  \Rightarrow
  \quad
  \mb{p}
  =
  \left(\begin{array}[c]{cccc}
   e^{-i\eta S} &                0 &  \dots &      0 \\
              0 & e^{-i\eta (S-1)} &  \dots &      0 \\
         \vdots &           \vdots & \ddots & \vdots \\
              0 &                0 &  \dots & e^{i\eta S}
  \end{array}\right)
  =
  e^{-i\eta \mb{S}_3}.
\end{gather}
Using this form of the matrix $\mb{p}$ we can then transform
eq~\eqref{eq:tildeHgS} as
\begin{align}
  \mb{p}^\dagger \widetilde{\mb{H}}_u^Z \mb{p}
  &=
  \frac{1}{2c} \widetilde{g}_{uv} e^{i\eta \mb{S}_3} \mb{S}_v e^{-i\eta \mb{S}_3},
  \\
  \label{eq:pHpS:3}
  \mb{p}^\dagger \widetilde{\mb{H}}_u^Z \mb{p}
  &=
  \frac{1}{2c} \widetilde{g}_{uv} \brr{e^{-i\eta \mb{R}_3}}_{vw} \mb{S}_w,
  \\
  \label{eq:pHpS:4}
  \mb{p}^\dagger \widetilde{\mb{H}}_u^Z \mb{p}
  &=
  \frac{1}{2c} \widebar{g}_{uv} \mb{S}_v,
  \quad
  \widebar{\bs{g}} = \widetilde{\bs{g}} e^{-i\eta \mb{R}_3},
\end{align}
where we have used identity~\eqref{SU:SO} derived in Appendix~\ref{app:rot}.
Because of the form of the matrix $\mb{R}_3$, eq~\eqref{R:def2}, the
exponential factor in eq~\eqref{eq:pHpS:4} can be written as follows
\begin{gather}
  \label{eq:Rtoeta}
  e^{-i\eta \mb{R}_3}
  = 
  \left(
    \begin{array}{cc|c}
        &         &  0  \\
      \multicolumn{2}{c|}{\smash{\raisebox{.5\normalbaselineskip}{$e^{-i\eta\sigma_2}$}}}
                  &  0  \\
      \hline \\[-\normalbaselineskip]
      0 &       0 &  1
    \end{array}
  \right)
  =
  \left(
    \begin{array}{cc|c}
        &         &  0  \\
      \multicolumn{2}{c|}{\smash{\raisebox{.5\normalbaselineskip}
                         {$\mr{cos}\,\eta-i\sigma_2\,\mr{sin}\,\eta$}}}
                  &  0  \\
      \hline \\[-\normalbaselineskip]
      0 &       0 &  1
    \end{array}
  \right)
  =
  \left(
    \begin{array}{ccc}
      \mr{cos}\,\eta & -\mr{sin}\,\eta &  0  \\
      \mr{sin}\,\eta &  \mr{cos}\,\eta &  0  \\
                 0 &             0 &  1
    \end{array}
  \right).
\end{gather}
Combining eqs~\eqref{eq:gtilde:almost}, \eqref{eq:pHpS:4} and~\eqref{eq:Rtoeta}
one gets
\begin{gather}
  \widebar{\bs{g}}
  = 
  \left(\begin{array}[c]{rrr}
   \widetilde{g}_{11} & \widetilde{g}_{12} &  0 \\
   \widetilde{g}_{21} & \widetilde{g}_{22} &  0 \\
   0 &  0 & \widetilde{g}_{33}
  \end{array}\right)
  \left(
    \begin{array}{ccc}
      \mr{cos}\,\eta & -\mr{sin}\,\eta &  0  \\
      \mr{sin}\,\eta &  \mr{cos}\,\eta &  0  \\
                 0 &             0 &  1
    \end{array}
  \right).
\end{gather}
Our goal is to diagonalize the matrix $\widebar{\bs{g}}$, a task which amounts
to solving the following two equations with unknown parameter $\eta$
\begin{align}
  \label{eq:v1rot}
  -\widetilde{g}_{11} \,\mr{sin}\,\eta + \widetilde{g}_{12} \,\mr{cos}\,\eta &= 0,
  \\
  \label{eq:v2rot}
  \widetilde{g}_{21} \,\mr{cos}\,\eta + \widetilde{g}_{22} \,\mr{sin}\,\eta &=0.
\end{align}
Because the rows of the $\widetilde{g}$-tensor are orthogonal vectors, see
eq~\eqref{eq:v1v2}, and the matrix $\mr{exp}(-i\eta \sigma_2)$ represents a
rotation in the two-dimensional real vector space, it is always possible to
find such a rotation if the vectors are nonzero.  Indeed, the solution of
eqs~\eqref{eq:v1rot} and~\eqref{eq:v2rot} reads
\begin{gather}
  \label{eq:eta}
  \eta
  = \mr{tan}^{-1}\brr{\frac{\widetilde{g}_{12}}{\widetilde{g}_{11}}}
  = \mr{tan}^{-1}\brr{-\frac{\widetilde{g}_{21}}{\widetilde{g}_{22}}}, 
\end{gather}
where we have utilized eq~\eqref{eq:v1v2}.  The parameter $\eta$ in
eq~\eqref{eq:eta} defines the matrix $\mb{p}$ in eqs~\eqref{eq:pahse:0}
and~\eqref{eq:pS3} that diagonalizes the $\widebar{g}$-tensor according to
eq~\eqref{eq:pHpS:4}.

The only loose end is the assumption that the rows of the
$\widetilde{g}$-tensor are nonzero. In the case of one or more zero rows, one
may fix the above procedure using a solution to analogous problem discussed in
section~\ref{sec:principal}.

{\bf The case of an arbitrary multiplicity:} In the procedure for the doublet
case we took advantage of the fact that, by fixing the determinant of the
matrix $\mb{C}$ to one, it will automatically belong to the $2$-dimensional
irreducible representation of SU$(2)$---{\it i.e.} the natural
representation---and can thus be expressed using the exponential mapping, see
eq~\eqref{eq:exp:U}. For the case of higher than doublet multiplicity, fixing
the determinant of $\mb{C}$ to one will make it a member of the group SU$(m)$,
but not necessarily a member of the $m$-dimensional irreducible representation
of SU$(2)$. However, as discussed earlier, there exists a choice of phase
factors $\mb{p}$ such that the matrix $\mb{C}\mb{p}$ is a member of the irrep
and thus can be parametrized according to eq~\eqref{eq:exp:U}. In addition, a
proper choice of $\mb{p}$ would lead to a diagonal g-tensor, {\it i.e.}
eq~\eqref{eq:HuZ:zz} would be satisfied ($u$ is not a summation index)
\begin{gather}
  \label{eq:HuZ:zz-1}
  \mb{p}^\dagger \widetilde{\mb{H}}_u^Z \mb{p}
  = \tfrac{1}{2c} \widetilde{g}_{uu} \mb{S}_u.
\end{gather}
Because there exist $\mb{p}$ for which this equation holds, the equation has
the following form for $u=1$ 
\begin{gather}
  \left(\begin{array}[c]{cccc}
   e^{-i\alpha_1} &              0 &  \dots &      0 \\
                0 & e^{-i\alpha_2} &  \dots &      0 \\
           \vdots &         \vdots & \ddots & \vdots \\
                0 &              0 &  \dots & e^{-i\alpha_m}
  \end{array}\right)
  \left(\begin{array}[c]{ccccc}
                    0 & r_1 e^{i\beta_1} &   \dots  &       0    &      0 \\
    r_1 e^{-i\beta_1} &                0 &   \dots  &       0    &      0 \\
               \vdots &           \vdots &  \ddots  &  \vdots    & \vdots \\
                    0 &                0 &   \dots  &       0    & r_{m-1}e^{i\beta_{m-1}} \\
                    0 &                0 &   \dots  & r_{m-1}e^{-i\beta_{m-1}} &         0 \\
  \end{array}\right)
  \times
  \\
  \times
  \left(\begin{array}[c]{cccc}
   e^{i\alpha_1} &             0 &  \dots &      0 \\
               0 & e^{i\alpha_2} &  \dots &      0 \\
          \vdots &        \vdots & \ddots & \vdots \\
               0 &             0 &  \dots & e^{i\alpha_m}
  \end{array}\right)
  = \tfrac{1}{2c} \widetilde{g}_{11}
  \left(\begin{array}[c]{ccccc}
          0 &    r_1 &   \dots  &       0 &      0 \\
        r_1 &      0 &   \dots  &       0 &      0 \\
     \vdots & \vdots &  \ddots  &  \vdots & \vdots \\
          0 &      0 &   \dots  &       0 & r_{m-1} \\
          0 &      0 &   \dots  & r_{m-1} &       0 \\
  \end{array}\right).
\end{gather}
Here only the upper- and lower-diagonal of the matrix $\widetilde{\mb{H}}_1^Z$
are nonzero. In addition, they are composed of complex numbers with arbitrary
phases and moduli equal to the off-diagonal elements of the matrix $\mb{S}_1$.
Because the beta parameters are known, one can write set of linear equations
for the unknown alpha parameters as
\begin{gather}
  \label{eq:get-pahses}
  \left(\begin{array}[c]{rrrrrr}
         1 &     -1 &      0 &  \dots &      0 &      0 \\
         0 &      1 &     -1 &  \dots &      0 &      0 \\
    \vdots \mkern 1.5mu & \vdots \mkern 1.5mu & \vdots \mkern 1.5mu & \ddots\, & \vdots \mkern 1.5mu & \vdots \mkern 1.5mu \\
         0 &      0 &      0 &  \dots &      1 &     -1 \\
         1 &      1 &      1 &  \dots &      1 &      1 \\
  \end{array}\right)
  \left(\begin{array}[c]{c}
    \alpha_1     \\
    \alpha_2     \\
    \vdots       \\
    \alpha_{m-1} \\
    \alpha_m     \\
  \end{array}\right)
  =
  \left(\begin{array}[c]{c}
    \beta_1     \\
    \beta_2     \\
    \vdots      \\
    \beta_{m-1} \\
    -\gamma     \\
  \end{array}\right).
\end{gather}
In eq~\eqref{eq:get-pahses} the last equation represents the fact that the new
coefficients $\widetilde{\mb{C}} = \mb{C} \mb{p}$ should belong to the
$m$-dimensional irreducible representation of SU$(m)$ with
$\mr{det}\,\widetilde{\mb{C}} = 1$.  However, the untransformed matrix $\mb{C}$
has a determinant with an arbitrary phase, $\mr{det}\,\mb{C} =
\mr{exp}(i\gamma)$, see also eqs~\eqref{eq:Cpahse1}--\eqref{eq:alpah:p}.  The
determinant of the matrix in eq~\eqref{eq:get-pahses} is nonzero, and therefore
the equation has a unique solution which defines the transformation matrix
$\mb{p}$, which in turn diagonalizes the g-tensor according to
eq~\eqref{eq:HuZ:zz-1}. Note that this procedure also works for the doublet
case.

The procedure for the diagonalization of the g-tensor under the assumptions
discussed in section~\ref{sec:introduction} can be summarized as follows:
\begin{enumerate}
\item
Choose the reference basis set $\{\Phi_n\}_{n=1}^m$ such that it satisfies the
same transformations under time-reversal symmetry as presented in
eq~\eqref{eq:TRrelations}, see section~\ref{sec:gt:construct}.
\item
Construct matrices $\mb{H}_u^Z$ and calculate the g-tensor, see
eq~\eqref{eq:HuZ}.
\item
Find the principal axes in real space, {\it i.e.} diagonalize the G-tensor, see
also eqs~\eqref{G:trans2} and~\eqref{G:diag1}.
\item 
Diagonalize the matrix $\mb{H}_3^Z$ and determine the $\widetilde{g}_{33}$
g-value, see eqs~\eqref{eq:H3Z}--\eqref{eigen:g33:2}.
\item
Calculate $\widetilde{\mb{H}}_1^Z$ and determine the phase factors (alpha
parameters) by solving the linear set of equation in eq~\eqref{eq:get-pahses}.
\item
Construct the matrix $\mb{p}$, $p_{kl} = \delta_{kl}\,\mr{exp}(i\alpha_k)$, and
diagonalize the g-tensor according to eq~\eqref{eq:HuZ:zz-1}.
\end{enumerate}

%
%
\begin{acknowledgement}
This project has been supported by the Slovak Grant Agencies VEGA and APVV
(contract no. 2/0135/21 and APVV-19-0516). In addition, this study has received
funding by the Operation Program of Integrated Infrastructure for the project,
UpScale of Comenius University Capacities and Competence in Research,
Development and Innovation,  ITMS2014+: 313021BUZ3, co-financed by the European
Regional Development Fund. I'm grateful to James R. Asher for thorough
proofreading of the text and valuable comments.
\end{acknowledgement}


\begin{appendices}

%
%
\section{}
\label{app:diag}
If one has access to the solutions of the perturbation-free eigenproblem,
eq~\eqref{eigen:H0}, it is convenient to express the wavefunctions $\Psi_k$ in
the truncated orthonormal basis of $\Phi_n$
\begin{gather}
  \Psi_k = C_{nk} \Phi_n.
\end{gather}
By projecting eq~\eqref{HPsiEPsi} onto the space defined by the orthonormal
wavefunctions $\Phi_n$, we then obtain the following eigenvalue matrix equation
\begin{gather}
  \big< \Phi_m \big| H^0 + H^Z \big| \Phi_n \big> \, C_{nk}
  =
  \mc{E}_k \, C_{mk},
  \\
  \left[ E_m \delta_{mn} + \big< \Phi_m \big| H^Z \big| \Phi_n \big> \right] \, C_{nk}
  =
  \mc{E}_k \, C_{mk}.
\end{gather}
Here we have used the eigenvalue expression in eq~\eqref{eigen:H0}, the
Kronecker delta function $\delta_{mn}$, and the fact that the set of basis
functions $\{\Phi_n\}$ is orthonormal, $\big< \Phi_m \big| \Phi_n \big> =
\delta_{mn}$.

%
%
\section{}
\label{app:odd-electrons}
 
{\bf Theorem:} For a system with an odd number of electrons described by a
time-reversal-symmetric Hamiltonian, every energy level is even-times
degenerate and the eigenfunctions can be chosen to form Kramers partners

\begin{gather}
  \label{eq:Nodd:KP}
  N_\mr{e} \text{ is odd},
  \quad [H,\mc{K}]=0,
  \quad H\Psi_i = E\Psi_i
  \\
  \Rightarrow
  \quad
  i=1 \dots 2N^E,
  \quad \{\Psi_i,\widebar{\Psi}_i\} \in \{\Psi_i\}_{i=1}^{2N^E}.
\end{gather}
In addition, $\{\Psi_i\}_{i=1}^{2N^E}$ forms an orthonormal set of wavefunctions.

\noindent{\bf Proof:} In Appendices~\ref{app:KP} and~\ref{app:kramers-partner}
it is shown that, for an odd number of electrons, Kramers partners are
orthonormal---{\it i.e.} they form two distinct wavefunctions---and that Kramers
partners are eigenfunctions belonging to the same energy level.  Therefore, one
can simply choose any eigenfunction $\Psi_1$, $H\Psi_1 = E\Psi_1$, and then by
forming its Kramers partner, $\widebar{\Psi}_1 = \mc{K}\Psi_1$, one gets another
eigenfunction with the same energy, $H\widebar{\Psi}_1 = E\widebar{\Psi}_1$.

Without loss of generality, we can then assume that a set of $2n$
eigenfunctions of $H$ are already orthonormal and form Kramers partners,
$\{\Psi_i,\widebar{\Psi}_i\}_{i=1}^{n}$. When starting the procedure outlined
below, simply set $n=1$. If the energy level is more than $2n$-times degenerate
one needs to construct a wavefunction $\Psi_{2n+1}$ that is orthonormal
to the set $\{\Psi_i,\widebar{\Psi}_i\}_{i=1}^{n}$ and that is an eigenfunction of
$H$ with the energy $E$.  One way to create $\Psi_{2n+1}$ is by the
following projection technique
\begin{gather}
  \label{eq:KP:tilde1}
  \Psi_{2n+1}
  =
  \brr{ 1 - \sum_{i=1}^{n}\left| \Psi_i \right> \left< \Psi_i \right| 
          - \sum_{i=1}^{n}\left| \widebar{\Psi}_i \right> \left< \widebar{\Psi}_i \right|} \Psi_j^\mr{orig}
  ,
  \\
  \label{eq:KP:tilde2}
  k = 1 \dots n
  \quad
  \Rightarrow
  \\
  \label{eq:KP:tilde3}
  \bapa{\Psi_k}{\Psi_{2n+1}}
  = 
  \bapa{\Psi_k}{\Psi_j^\mr{orig}} 
  -
  \sum_{i=1}^n \bapa{\Psi_k}{\Psi_i} \bapa{\Psi_i}{\Psi_j^\mr{orig}}
  -
  \sum_{i=1}^n \bapa{\Psi_k}{\widebar{\Psi}_i} \bapa{\widebar{\Psi}_i}{\Psi_j^\mr{orig}}
  = 0
  ,
  \\
  \label{eq:KP:tilde4}
  \bapa{\widebar{\Psi}_k}{\Psi_{2n+1}}
  = 
  \bapa{\widebar{\Psi}_k}{\Psi_j^\mr{orig}} 
  -
  \sum_{i=1}^n \bapa{\widebar{\Psi}_k}{\Psi_i} \bapa{\Psi_i}{\Psi_j^\mr{orig}}
  -
  \sum_{i=1}^n \bapa{\widebar{\Psi}_k}{\widebar{\Psi}_i} \bapa{\widebar{\Psi}_i}{\Psi_j^\mr{orig}}
  = 0
  .
\end{gather}
where we have utilized the fact that $\bapa{\Psi_k}{\Psi_i}
=\bapa{\widebar{\Psi}_k}{\widebar{\Psi}_i} = \delta_{ki}$ and
$\bapa{\Psi_k}{\widebar{\Psi}_i} = \bapa{\widebar{\Psi}_k}{\Psi_i} = 0$.  The
wavefunction $\Psi_j^\mr{orig}$ is from the initial set of eigenfunctions of
the Hamiltonian---{\it i.e.} a set that does not necessarily consist of sets of
Kramers partners. If $\Psi_j^\mr{orig}$ belongs to the subspace formed by
$\{\Psi_i,\widebar{\Psi}_i\}_{i=1}^{n}$ then $\Psi_{2n+1}$ is zero, in which
case one needs to repeat the procedure with a different $\Psi^\mr{orig}$ until
a nonzero wavefunction is obtained.  When successful one can then normalize
the wavefunction $\Psi_{2n+1}$. Because $\Psi_{2n+1}$ is a linear combination
of eigenfunctions of the same energy level, it is also an eigenfunction with
that same energy as well.

According to Appendix~\ref{app:KP} the wavefunction $\widebar{\Psi}_{2n+1}$ is
orthogonal to $\Psi_{2n+1}$, and because $\Psi_{2n+1}$ is orthonormal to the
set $\{\Psi_i,\widebar{\Psi}_i\}_{i=1}^{n}$,
eqs~\eqref{eq:KP:tilde2}--\eqref{eq:KP:tilde4}, $\widebar{\Psi}_{2n+1}$ is
orthonormal to this set as well ($k = 1 \dots n$)
\begin{gather}
  \bapa{\widebar{\Psi}_{2n+1}}{\Psi_k}
  = \bapa{\mc{K}\Psi_{2n+1}}{\Psi_k}
  = \bapa{\Psi_{2n+1}}{\mc{K}^\dagger\Psi_k}^\ast
  = -\bapa{\Psi_{2n+1}}{\widebar{\Psi}_k}^\ast
  = 0,
  \\
  \bapa{\widebar{\Psi}_{2n+1}}{\widebar{\Psi}_k}
  = \bapa{\mc{K}\Psi_{2n+1}}{\widebar{\Psi}_k}
  = \bapa{\Psi_{2n+1}}{\mc{K}^\dagger\mc{K}\Psi_k}^\ast
  = \bapa{\Psi_{2n+1}}{\Psi_k}^\ast
  = 0,
\end{gather}
where we have utilized the fact that for an odd number of electrons the
following holds [see also eq~\eqref{eq:KK}]
\begin{align}
  \mc{K}\mc{K} &= -1,
  \\
  \mc{K}^\dagger\mc{K}\mc{K} &= -\mc{K}^\dagger,
  \\
  \mc{K}^\dagger &= -\mc{K},
\end{align}
and that $\mc{K}$ is a unitary operator, $\mc{K}^\dagger\mc{K}=1$.  In
addition, note that Kramers partners have the same norm, see
Appendix~\eqref{app:kramers-partner}, and thus if $\Psi_{2n+1}$ is normalized
to one then $\widebar{\Psi}_{2n+1}$ is normalized to one as well. If $2n+2 <
2N^E$ then one should repeat the described procedure with $n=n+1$. Finally, it
is now clear that every energy level must be even-times degenerate, because
when using the procedure described above one adds eigenfunctions in pairs. QED.

%
%
\section{}
\label{app:non-kramers-partner}

{\bf Theorem:} The non-Kramers partners\cite{Griffith:1963:non-KP} can be
constructed from the non-magnetic wavefunctions as follows
\begin{gather}
  \label{nonKP}
  \begin{array}[c]{cc}
   \mathcal{K}\Psi_1 = \Psi_1 \\
   \mathcal{K}\Psi_2 = \Psi_2
  \end{array}
  \quad
  \Rightarrow
  \quad
  \begin{array}[c]{cc}
   \Phi = \tfrac{1}{\sqrt{2}} \brr{ \Psi_1 + i \Psi_2 } \\
   \widebar\Phi = \tfrac{1}{\sqrt{2}} \brr{ \Psi_1 - i \Psi_2 } \\
  \end{array}
  ,
  \\
  \bapa{\Psi_i}{\Psi_j} = \delta_{ij},
  \quad i,j = 1,2
  \quad
  \Rightarrow
  \quad
  \bapa{\Phi}{\Phi} = \bapa{\widebar\Phi}{\widebar\Phi} = 1.
\end{gather}

\noindent{\bf Proof:} Using the fact that $\mc{K}$ is an antilinear operator
[see eq~\eqref{antilinear2}], it holds that
\begin{gather}
  \widebar{\Phi} = \mc{K}\Phi = \tfrac{1}{\sqrt{2}} \mc{K}\brr{ \Psi_1 + i \Psi_2 }
  = \tfrac{1}{\sqrt{2}} \brr{ \mc{K}\Psi_1 - i \mc{K}\Psi_2 }
  = \tfrac{1}{\sqrt{2}} \brr{ \Psi_1 - i \Psi_2 }
  \\
  \bapa{\Phi}{\Phi}
  = \tfrac{1}{2}\bapa{\Psi_1 + i \Psi_2}{\Psi_1 + i \Psi_2}
  \\
  = \tfrac{1}{2}\brs{ \bapa{\Psi_1}{\Psi_1}
                      +\bapa{\Psi_2}{\Psi_2}
                      +2\mc{R}\brr{ i\bapa{\Psi_1}{\Psi_2} } }
  = 1.
\end{gather}
And according to Appendix~\ref{app:kramers-partner} (non-)Kramers partners have
the same norm. QED.

%
%
\section{}
\label{app:non-magnetic1}
 
{\bf Theorem:} Eigenfunctions of the time-reversal symmetric Hamiltonian for a
system with an even number of electrons can be chosen such that they are invariant
under time-reversal, {\it i.e.} they are non-magnetic (see
Appendix~\ref{app:non-magnetic2})
\begin{equation}
  \label{eq:Neven:non-magnetic}
  N_\mr{e} \text{ is even},
  \quad H\Psi = E\Psi,
  \quad [H,\mc{K}]=0
  \quad
  \Rightarrow
  \quad
  \mathcal{K}\Psi = \Psi,
\end{equation}
and form an orthonormal set of wavefunctions.

\noindent{\bf Proof:} To prove the above theorem it is sufficient to show that
eq~\eqref{eq:Neven:non-magnetic} holds for an energy level of arbitrary
degeneracy, $i=1 \dots N^E$ 
\begin{equation}
  \label{eq:Neven:non-magnetic-1}
  N_\mr{e} \text{ is even},
  \quad H\Psi_i = E\Psi_i,
  \quad [H,\mc{K}]=0
  \quad
  \Rightarrow
  \quad
  \mathcal{K}\Psi_i = \Psi_i.
\end{equation}
Without loss of generality, let us assume that a set of $n-1$ eigenfunctions of
$H$ are already orthonormal and non-magnetic, $\mathcal{K}\Psi_i^\mr{nm} =
\Psi_i^\mr{nm}$ for $i=1 \dots n-1$, and that the $n$th eigenfunction
$\Psi_{n}$ is orthogonal to this set and normalized to one. When starting the
procedure outlined below simply set $n=1$.  The Kramers partner
$\widebar{\Psi}_n \coloneqq \mc{K}\Psi_n$ is the eigenfunction of $H$ with the
energy $E$ and has the same norm as $\Psi_n$, see
Appendix~\ref{app:kramers-partner}. The non-magnetic eigenfunction of $H$ with
energy $E$ can be constructed as
\begin{gather}
  \label{eq:nm-construction}
  \Psi_n^\mr{nm} = c_n\Psi_n + c_n^\ast\widebar{\Psi}_n,
\end{gather}
where $c_n=r_ne^{i\alpha_n}$ is a complex number chosen such that
$\Psi_n^\mr{nm}$ is normalized to one
\begin{gather}
  \bapa{\Psi_n^\mr{nm}}{\Psi_n^\mr{nm}}
  =
  |c_n|^2 \bapa{\Psi_n}{\Psi_n} +
  |c_n|^2 \bapa{\widebar{\Psi}_n}{\widebar{\Psi}_n} +
  c_n^2 \bapa{\widebar{\Psi}_n}{\Psi_n} +
  (c_n^2)^\ast \bapa{\Psi_n}{\widebar{\Psi}_n}
  \\
  =
  2|c_n|^2 + 2 \mc{R} \brr{c_n^2 \bapa{\widebar{\Psi}_n}{\Psi_n}} =
  2 r_n^2 \brs{1 + \mc{R} \brr{e^{i2\alpha_n} \bapa{\widebar{\Psi}_n}{\Psi_n}} }
  \overset{!}{=} 1.
\end{gather}
One solution for the parameters $r_n$ and $\alpha_n$ can be expressed as
\begin{gather}
  \label{eq:c:aa0}
  \bapa{\widebar{\Psi}_n}{\Psi_n} = 0
  \quad
  \Rightarrow
  \quad
  r_n = \tfrac{1}{\sqrt{2}}, \quad \forall \alpha_n \in \mathbb{R},
  \\
  \label{eq:c:aa1}
  \bapa{\widebar{\Psi}_n}{\Psi_n} \ne 0
  \quad
  \Rightarrow
  \quad
  r_n = \brs{ 2 \brr{1 + \brp{\bapa{\widebar{\Psi}_n}{\Psi_n}}}}^{-\tfrac{1}{2}} , \quad
  \alpha_n = - \tfrac{1}{2}\,\mr{arccos}
           \brr{ \frac{\mc{R}\bapa{\widebar{\Psi}_n}{\Psi_n}}{\brp{\bapa{\widebar{\Psi}_n}{\Psi_n}}} }.
\end{gather}
The wavefunction $\Psi_n^\mr{nm}$ is non-magnetic
\begin{gather}
  \label{eq:nm-proof}
  \mc{K}\Psi_n^\mr{nm} = \mc{K}c_n\Psi_n + \mc{K}c_n^\ast\widebar{\Psi}_n
  = c_n^\ast\mc{K}\Psi_n + c_n\mc{K}\widebar{\Psi}_n
  = c_n^\ast\widebar{\Psi}_n + c_n\Psi_n = \Psi_n^\mr{nm}.
\end{gather}
Furthermore, $\Psi_n^\mr{nm}$ is an eigenfunction of $H$ with eigenvalue $E$,
because it is composed of two wavefunctions $\Psi_n$ and $\widebar{\Psi}_n$
which are also eigenfunctions of $H$ with eigenvalue $E$ [see also
~\eqref{eq:psibar4}].  In eq~\eqref{eq:nm-proof} we have utilized the relation
$\mc{K}^2=1$, which holds for systems with even numbers of electrons, see
eq~\eqref{eq:KK}. Finally, because $\Psi_n$ is orthogonal to the set of
wavefunctions $\{\Psi_i^\mr{nm}\}_{i=1}^{n-1}$,
$\bapa{\Psi_i^\mr{nm}}{\Psi_n}=0$, the non-magnetic eigenfunction
$\Psi^\mr{nm}_n$ is orthogonal to this set as well
\begin{gather}
  \bapa{\Psi_i^\mr{nm}}{\Psi_n^\mr{nm}} = c_n \bapa{\Psi_i^\mr{nm}}{\Psi_n}
                                        + c_n^\ast \bapa{\Psi_i^\mr{nm}}{\widebar{\Psi}_n}
  = c_n^\ast \bapa{\Psi_i^\mr{nm}}{\mc{K}^\dagger\mc{K}\widebar{\Psi}_n}
  \\
  = c_n^\ast \bapa{\mc{K} \Psi_i^\mr{nm}}{\mc{K}\mc{K}\Psi_n}^\ast
  = c_n^\ast \bapa{\Psi_i^\mr{nm}}{\Psi_n}^\ast = 0.
\end{gather}

To repeat the procedure outlined above, one first needs to construct the
wavefunction $\Psi_{n+1}$ that is orthonormal to our given set of non-magnetic
orthonormal eigenfunctions $\{\Psi_i^\mr{nm}\}_{i=1}^{n}$ and that is the
eigenfunction of $H$ with the energy $E$.  One way to construct such
wavefunction is to project out every wavefunction
$\{\Psi_i^\mr{nm}\}_{i=1}^{n}$ from the eigenfunction $\Psi_j^\mr{orig}$
\begin{gather}
  \label{eq:psi:tilde1}
  \Psi_{n+1}
  =
  \brr{ 1 - \sum_{i=1}^n\left| \Psi_i^\mr{nm} \right> \left< \Psi_i^\mr{nm} \right|} \Psi_j^\mr{orig}
  ,
  \\
  \label{eq:psi:tilde2}
  k = 1 \dots n
  \quad
  \Rightarrow
  \quad
  \bapa{\Psi_k^\mr{nm}}{\Psi_{n+1}}
  = 
  \bapa{\Psi_k^\mr{nm}}{\Psi_j^\mr{orig}} 
  -
  \sum_{i=1}^n \bapa{\Psi_k^\mr{nm}}{\Psi_i^\mr{nm}} \bapa{\Psi_i^\mr{nm}}{\Psi_j^\mr{orig}}
  = 0
  .
\end{gather}
where $\bapa{\Psi_k^\mr{nm}}{\Psi_i^\mr{nm}} = \delta_{ki}$ and
$\Psi_j^\mr{orig}$ is an eigenfunction from the starting set of wavefunctions
in eq~\eqref{eq:Neven:non-magnetic-1}. However, if $\Psi_j^\mr{orig}$ belongs
to the subspace represented by the wavefunctions
$\{\Psi_i^\mr{nm}\}_{i=1}^{n}$, eq~\eqref{eq:psi:tilde1} gives a zero
wavefunction. In this case one just needs to chose a different $\Psi^\mr{orig}$
and repeat until a nonzero wavefunction is obtained.  Finally, normalize
the nonzero $\Psi_{n+1}$ and start the process from the beginning with $n=n+1$.
QED.

%
%
\section{}
\label{app:non-magnetic2}

{\bf Theorem:} For operators that are time-reversal-antisymmetric and
Hermitian, the non-magnetic states have vanishing expectation values
\begin{equation}
  \label{eq:non-magnetic}
  O^\dagger = O,
  \quad
  \mc{K} O \mc{K}^\dagger = -O,
  \quad
  \mc{K}\Psi = \Psi
  \quad
  \Rightarrow
  \quad
  \bappa{\Psi}{O}{\Psi} = 0.
\end{equation}

\noindent{\bf Proof:} Utilizing the fact that the time-reversal operator is
unitary, eq~\eqref{antilinear1}, a proper definition of the adjoint of the
antilinear operator, eq~\eqref{antilinear3}, and the fact that the expectation
values of a Hermitian operator are real, eq~\eqref{eq:hermitian}, one can show
\begin{gather}
  \bappa{\Psi}{O}{\Psi}
  = \bappa{\Psi}{\mc{K}^\dagger\mc{K}O\mc{K}^\dagger\mc{K}}{\Psi}
  = - \bappa{\Psi}{\mc{K}^\dagger O\mc{K}}{\Psi}
  = - \bappa{\mc{K}\Psi}{O}{\mc{K}\Psi}^\ast
  \\
  = - \bappa{\Psi}{O}{\Psi}^\ast
  = - \bappa{\Psi}{O}{\Psi}.
\end{gather}
The only number that can be positive and negative at the same time is zero.
QED. Note that the reason that the wavefunctions for which $\widebar\Psi =
\Psi$ are called "non-magnetic" is that all operators which represent magnetic
interactions are time-reversal-antisymmetric and Hermitian; and thus these
wavefunctions have zero expectation values with magnetic operators according
to theorem~\eqref{eq:non-magnetic}.

%
%
\section{}
\label{app:nonKramers-off}
\noindent{\bf Theorem}: For non-Kramers partners the off-diagonal elements of
the time-reversal-anti\-symmetric and Hermitian operator $O$ are zero
\begin{equation}
  O^\dagger = O,
  \quad
  O\mc{K} = -\mc{K}O,
  \quad
  \widebar\Psi = \mc{K} \Psi,
  \quad
  \mc{K}^2 = 1
  \quad
  \Rightarrow
  \quad
  \bappa{\Psi}{O}{\widebar\Psi} = 0
\end{equation}

\noindent{\bf Proof}: Non-Kramers pairs (see
Appendix~\ref{app:non-kramers-partner}) may arise only in the case of a system
with an even number of electrons, because they are constructed from non-magnetic
wavefunctions, $\widebar\psi = \psi$, which cannot occur for odd $N_\mr{e}$,
where $\bapa{\psi}{\widebar\psi}=0$ (see Appendix~\ref{app:KP}). For the case
of a system with an even number of electrons it holds that $\mc{K}^2 = 1$ [see
eq~\eqref{eq:KK}], and we can write
\begin{align}
  \bappa{\Psi}{O}{\widebar\Psi}
  &= \bappa{\Psi}{O\mc{K}}{\Psi}
  = - \bappa{\Psi}{\mc{K} O}{\Psi}
  = - \bappa{\mc{K}^\dagger\Psi}{O}{\Psi}^\ast
  = - \bappa{\mc{K}\Psi}{O}{\Psi}^\ast
  \\
  &= - \bappa{\widebar\Psi}{O}{\Psi}^\ast
  = - \bapa{O\Psi}{\widebar\Psi}
  = - \bappa{\Psi}{O^\dagger}{\widebar\Psi}
  = - \bappa{\Psi}{O}{\widebar\Psi}.
\end{align}
Here we are employing an identity valid for even $N_\mr{e}$
\begin{align}
  \mc{K}^2 &= 1,
  \\
  \mc{K}^2\mc{K}^\dagger &= \mc{K}^\dagger,
  \\
  \mc{K} &= \mc{K}^\dagger,
\end{align}
as well as using the properties of the time-reversal operator in
eqs~\eqref{antilinear1} and~\eqref{antilinear3}, the definition of a Hermitian
operator, $O^\dagger = O$, and the definition of the adjoint of a linear
operator in eq~\eqref{eq:linear-adjoint}. Finally, one must note that zero is
the only number that can be simultaneously both positive and negative. QED.

%
%
\section{}
\label{app:K}

In the four-component framework the many-electron time-reversal operator, for
the system with $N_\mr{e}$ electrons, has the form 
\begin{gather} \label{def:K}
  \mc{K} = \prod_{i=1}^{N_\mr{e}} K_i,
  \qquad
  K_i = -i \Sigma_{y,i} K_{0,i},
  \qquad
  \Sigma_{y,i} =
  \left(
  \begin{matrix}
      \sigma_{y,i}  &  0  \\
      0         &  \sigma_{y,i}
  \end{matrix}
  \right),
\end{gather}
where $K_i$ is the one-electron time-reversal operator, $K_{0,i}$ represents
the complex conjugation operator, $K_{0,i}\,\phi(\vec{r}_i) =
\phi^\ast(\vec{r}_i)$, $\sigma_{y,i}$ is the $y$th Pauli matrix, and index $i$
indicates on which electron the operators are acting. Because $\mc{K}$ and
$K_i$ are antilinear unitary operators it holds that
\begin{gather}
  \mc{K}^\dagger \mc{K} = \mc{K} \mc{K}^\dagger = 1,
  \label{antilinear1}
  \\
  \mc{K} \left( a\Psi + b\Phi \right) = a^\ast \mc{K}\Psi + b^\ast \mc{K}\Phi,
  \label{antilinear2}
  \\
  \big< \Psi \big| \mc{K}\Phi \big> = \big< \mc{K}^\dagger \Psi \big| \Phi \big>^\ast,
  \label{antilinear3}
  \\
  \nonumber
  \\
  K_i^\dagger K_i = K_i K_i^\dagger = 1,
  \label{antilinear4}
  \\
  K_i \left( a\psi + b\phi \right) = a^\ast K_i\psi + b^\ast K_i\phi,
  \label{antilinear5}
  \\
  \big< \psi \big| K_i\phi \big> = \big< K_i^\dagger \psi \big| \phi \big>^\ast,
  \label{antilinear6}
\end{gather}
where $a,b \in \mathbb C$; $\psi, \phi \in [L^2(\mathbb R^3)]^4$ are
four-spinors (complex vector-valued square-integrable functions from
$\mathbb{R}^3$ to $\mathbb{C}^4$); and $\Psi,\Phi \in V^{N_\mr{e}}$, with
$V^{N_\mr{e}} = S^- H^{\otimes N_\mr{e}}$ and $H=[L^2(\mathbb R^3)]^4$, are
many-electron wavefunctions that belong to the Fock subspace for $N_\mr{e}$
fermions. Because the Fock subspace $V^{N_\mr{e}}$ is equipped with the inner
product
\begin{equation}
  \label{eq:ip-prop-0}
  \big<.\big|.\big> : \quad V^{N_\mr{e}} \times V^{N_\mr{e}} \rightarrow \mathbb{C},
\end{equation}
defined by the properties
\begin{gather}
  \label{eq:ip-prop-1}
  \bapa{\Psi}{\Phi}^\ast = \bapa{\Phi}{\Psi},
  \\
  \label{eq:ip-prop-2}
  \bapa{a\Psi_1 + b\Psi_2}{\Phi} = a \bapa{\Psi_1}{\Phi} + b \bapa{\Psi_2}{\Phi},
  \quad
  a,b \in \mathbb{C},
  \\
  \label{eq:ip-prop-3}
  \Psi \ne 0 
  \quad
  \Rightarrow
  \quad
  \bapa{\Psi}{\Psi} > 0,
\end{gather}
it forms a Hilbert space. Finally, note that the definition of the adjoint of
the linear operator $O$---in contrast to the antilinear operator above---is
written as
\begin{gather}
  \label{eq:linear-adjoint}
  \bapa{\Psi}{O\Phi} = \bapa{O^\dagger\Psi}{\Phi}.
\end{gather}
For a more detailed description of antilinear operators, we refer the
interested reader to the excellent book by Messiah.\cite{Messiah-book} Note
especially the definition of the adjoint of the antilinear operator, see
eq~\eqref{antilinear3} and related discussion in Chapter XV in
ref~\citenum{Messiah-book}.

%
%
\section{}
\label{app:Kramers-spin}
\noindent{\bf Theorem}:
For a time-reversal-antisymmetric and Hermitian operator $O$ the expectation
values of (non-)Kramers partners have opposite sign
\begin{equation}
  O^\dagger = O,
  \quad
  \mc{K}^\dagger O\mc{K} = -O,
  \quad
  \widebar\Psi = \mc{K} \Psi
  \quad  
  \Rightarrow
  \quad
  \bappa{\widebar\Psi}{O}{\widebar\Psi} = -\bappa{\Psi}{O}{\Psi}.
\end{equation}

\noindent{\bf Proof}:
Because $O$ is a Hermitian operator, $O^\dagger = O$, the  expectation values
of any wavefunction are real
\begin{equation}
  \label{eq:hermitian}
  \bappa{\Psi}{O}{\Psi}^\ast
  = \bapa{O\Psi}{\Psi}
  = \bappa{\Psi}{O^\dagger}{\Psi}
  = \bappa{\Psi}{O}{\Psi}.
\end{equation}
Considering eq~\eqref{eq:hermitian}, the properties of the time-reversal
operator $\mc{K}$, eqs~\eqref{antilinear1}--\eqref{antilinear3}, and the fact
that $O$ is time-reversal-antisymmetric, $\mc{K}^\dagger O \mc{K} = - O$, one
can write
\begin{align}
  \big< \widebar\Psi \big| O \big| \widebar\Psi \big>
  &=
  \big< \mc{K} \Psi \big| O \big| \mc{K} \Psi \big>
  =
  \big< \Psi \big| \mc{K}^\dagger O \mc{K} \big| \Psi \big>^\ast
  \\
  &=
  - \big< \Psi \big| O \big| \Psi \big>^\ast
  =
  - \big< \Psi \big| O \big| \Psi \big>.
\end{align}
QED.

%
%
\section{}
\label{app:phase}

\noindent{\bf Theorem}: For the doublet case, if one changes the phase of the
Kramers pair $\{\Psi, \widebar\Psi\}$ associated with the $g$-tensor
\begin{align}
  \label{gt:2-psi}
  \begin{array}[c]{rr}
  g_{u1} =& 4c\,\mc{R} \bappa{\Psi}{H^Z_u}{\widebar\Psi},
  \\
  g_{u2} =&-4c\,\mc{I} \bappa{\Psi}{H^Z_u}{\widebar\Psi},
  \\
  g_{u3} =& 4c\,\bappa{\Psi}{H^Z_u}{\Psi},
  \end{array}
\end{align}
the new $\widetilde{g}$-tensor of the phase-shifted Kramers pair
$\{\widetilde{\Psi}, \widebar{\widetilde{\Psi}}\}$
\begin{align}
  \label{gt:2-phi}
  \begin{array}[c]{rr}
  \widetilde{g}_{u1} =& 4c\,\mc{R} \bappa{\widetilde{\Psi}}{H^Z_u}{\widebar{\widetilde{\Psi}}},
  \\
  \widetilde{g}_{u2} =&-4c\,\mc{I} \bappa{\widetilde{\Psi}}{H^Z_u}{\widebar{\widetilde{\Psi}}},
  \\
  \widetilde{g}_{u3} =& 4c\,\bappa{\widetilde{\Psi}}{H^Z_u}{\widetilde{\Psi}},
  \end{array}
\end{align}
can be written in the form
\begin{gather}
  \widetilde{\Psi} = e^{i\alpha} \Psi,
  \quad
  \alpha \in \mathbb{R}
  \\
  \label{eq:gtogbyphase}
  \Rightarrow
  \quad
  \left(\begin{array}[c]{rrr}
   \widetilde{g}_{11} & \widetilde{g}_{12} & \widetilde{g}_{13} \\
   \widetilde{g}_{21} & \widetilde{g}_{22} & \widetilde{g}_{23} \\
   \widetilde{g}_{31} & \widetilde{g}_{32} & \widetilde{g}_{33}
  \end{array}\right)
  =
  \left(\begin{array}[c]{rrr}
   g_{11} & g_{12} & g_{13} \\
   g_{21} & g_{22} & g_{23} \\
   g_{31} & g_{32} & g_{33}
  \end{array}\right)
  \left(
    \begin{array}{ccc}
      \mr{cos}\brr{2\alpha} &  \mr{sin}\brr{2\alpha} &  0  \\
     -\mr{sin}\brr{2\alpha} &  \mr{cos}\brr{2\alpha} &  0  \\
                 0 &             0 &  1
    \end{array}
  \right).
\end{gather}

\noindent{\bf Proof}: Because the time-reversal operator is antilinear, see
eq~\eqref{antilinear2}, the Kramers partner of $\widetilde{\Psi}$ transforms as
follows
\begin{equation}
  \label{eq:change-phase}
  \widetilde{\Psi} = e^{i\alpha} \Psi
  \quad
  \Rightarrow
  \quad
  \widebar{\widetilde{\Psi}} = e^{-i\alpha} \widebar\Psi.
\end{equation}
Inserting the relations between $\{\widetilde{\Psi}, \widebar{\widetilde{\Psi}}\}$
and $\{\Psi, \widebar\Psi\}$ into eqs~\eqref{gt:2-phi}, employing the identity
\begin{equation}
  e^{i\alpha} = \mr{cos}\,\alpha + i\,\mr{sin}\,\alpha,
\end{equation}
and using the relations for the $g$-tensor in eq~\eqref{gt:2-psi}, one can expand
the components of the $\widetilde{g}$-tensor in terms of the elements of the g-tensor
\begin{gather}
  \label{eq:phase-gtilde-x}
  \begin{array}[c]{c}
    \widetilde{g}_{u1}
    = 4c\,\mc{R} \bappa{e^{i\alpha} \Psi}{H^Z_u}{e^{-i\alpha} \widebar\Psi}
    = 4c\,\mc{R} \brc{ e^{-i2\alpha} \bappa{\Psi}{H^Z_u}{\widebar\Psi} }
    \\
    = 4c\,\mc{R} \brc{ \big[ \mr{cos}\brr{2\alpha} - i\,\mr{sin}\brr{2\alpha} \big]
                       \brs{ \mc{R}\bappa{\Psi}{H^Z_u}{\widebar\Psi}
                             + i \, \mc{I}\bappa{\Psi}{H^Z_u}{\widebar\Psi} }
                     }
    \\
    =   4c \: \mr{cos}\brr{2\alpha} \mc{R}\bappa{\Psi}{H^Z_u}{\widebar\Psi}
      + 4c \: \mr{sin}\brr{2\alpha} \mc{I}\bappa{\Psi}{H^Z_u}{\widebar\Psi}
    \\
    = \mr{cos}\brr{2\alpha} g_{u1} - \mr{sin}\brr{2\alpha} g_{u2},
  \end{array}
  \\
  \label{eq:phase-gtilde-y}
  \begin{array}[c]{c}
    \widetilde{g}_{u2}
    = - 4c\,\mc{I} \bappa{e^{i\alpha} \Psi}{H^Z_u}{e^{-i\alpha} \widebar\Psi}
    = - 4c\,\mc{I} \brc{ e^{-i2\alpha} \bappa{\Psi}{H^Z_u}{\widebar\Psi} }
    \\
    = - 4c\,\mc{I} \brc{ \big[ \mr{cos}\brr{2\alpha} - i\,\mr{sin}\brr{2\alpha} \big]
                         \brs{ \mc{R}\bappa{\Psi}{H^Z_u}{\widebar\Psi}
                               + i \, \mc{I}\bappa{\Psi}{H^Z_u}{\widebar\Psi} }
                       }
    \\
    = - 4c \: \mr{cos}\brr{2\alpha} \mc{I}\bappa{\Psi}{H^Z_u}{\widebar\Psi}
      + 4c \: \mr{sin}\brr{2\alpha} \mc{R}\bappa{\Psi}{H^Z_u}{\widebar\Psi}
    \\
    = \mr{cos}\brr{2\alpha} g_{u2} + \mr{sin}\brr{2\alpha} g_{u1},
  \end{array}
  \\
  \label{eq:phase-gtilde-z}
  \widetilde{g}_{u3}
  = 4c \bappa{e^{i\alpha} \Psi}{H^Z_u}{e^{i\alpha} \Psi}
  = 4c \bappa{\Psi}{H^Z_u}{\Psi}
  = g_{u3}.
\end{gather}
Eqs~\eqref{eq:phase-gtilde-x}--\eqref{eq:phase-gtilde-z} can be written in the
matrix form presented in eq~\eqref{eq:gtogbyphase}. QED.

Note that this theorem is a special case of the transformation of the g-tensor
presented in eqs~\eqref{eq:tildeg:UpSUp}--\eqref{eq:tildeg:OpS1}, where the key
identity, eq~\eqref{SU:SO}, has---from the point of view of this appendix---the
form
\begin{align}
  e^{-i\alpha \bs{\sigma}_z} \, \bs{\sigma}_u \, e^{i\alpha \bs{\sigma}_z}
  =
  \left( e^{i 2 \alpha \mb{R}_z} \right)_{uv} \bs{\sigma}_v,
\end{align}
with $\theta = -2\alpha$ and $\vec{n} = (0,0,1)$. The transformation of the basis
wavefunctions can be written in the matrix form
\begin{equation}
  e^{i\alpha \bs{\sigma}_z}
  =
  \left(\begin{array}[c]{cc}
   e^{i\alpha} & 0 \\
             0 & e^{-i\alpha}
  \end{array}\right), 
\end{equation}
which corresponds to a change of the phase in eq~\eqref{eq:change-phase},
and according to eq~\eqref{eq:Rtoeta} the three-dimensional rotation
$\mr{exp}\,(i2\alpha\mb{R}_z)$ has the form of the transformation matrix in
eq~\eqref{eq:gtogbyphase}.

%
%
\section{}
\label{app:KP}

{\bf Theorem:} For a system with an odd number of electrons described by the
time-reversal-symmetric Hamiltonian, every energy level is at least twofold
degenerate and is formed by Kramers partners.

\noindent{\bf Proof:} First, let us apply the time-reversal operator
$\mc{K}$ on the eigenvalue equation \eqref{eigen:H0}
\begin{gather}
  H^0 \Phi_k = E_k \Phi_k,
  \\
  \mc{K} H^0 \mc{K}^\dagger \mc{K} \Phi_k = E_k \mc{K} \Phi_k.
\end{gather}
Because the Hamiltonian $H^0$ is time-reversal-symmetric, $\mc{K} H^0
\mc{K}^\dagger = H^0$, we can write
\begin{gather}
  H^0 \Phi_k = E_k \Phi_k,
  \label{KP1}
  \\
  H^0 \widebar\Phi_k = E_k \widebar\Phi_k,
  \label{KP2}
\end{gather}
where we have utilized the definition of the Kramers partner, $\widebar\Phi_k =
\mc{K} \Phi_k$. From expressions \eqref{KP1} and \eqref{KP2} it follows that
wavefunctions $\Phi_k$ and $\widebar\Phi_k$ share the same energy. To prove
that the energy level $E_k$ is at least doubly-degenerate we need to show that
the wavefunctions are different. To do so we will show that they are
orthogonal. Note that two wavefunctions are the same, up to a phase factor, if
their inner product is a phase factor.  For the inner product of the above
Kramers pair one can write
\begin{align}
  \big< \Phi_k \big| \widebar\Phi_k \big>
  &=
  \big< \mc{K}^\dagger \mc{K} \Phi_k \big| \mc{K}\Phi_k \big>
  =
  \big< \mc{K} \Phi_k \big| \mc{K} \mc{K}\Phi_k \big>^\ast
  \nonumber
  \\
  &=
  \big< \mc{K} \mc{K} \Phi_k \big| \widebar\Phi_k \big>
  =
  \left( -1 \right)^{N_\mr{e}}
  \big< \Phi_k \big| \widebar\Phi_k \big>.
  \label{orthogonal}
\end{align}
Here we have utilized the fact that $\mc{K}$ is the antilinear unitary operator,
\eqref{antilinear1} and \eqref{antilinear3}, and the following identity for the
square of the many-electron time-reversal operator
\begin{gather}
  \label{eq:KK}
  \mc{K} \mc{K}
  =
  \left( \prod_i^{N_\mr{e}} K_i \right) \left( \prod_j^{N_\mr{e}} K_j \right)
  =
  \prod_i^{N_\mr{e}} K_i K_i
  =
  \prod_i^{N_\mr{e}} \left(-1\right)
  =
  \left(-1\right)^{N_\mr{e}},
\end{gather}
where $(K_i)^2=-1$ follows directly from the definition of $K_i$
in eq~\eqref{def:K}.

From eq~\eqref{orthogonal} it follows that for an odd number of electrons the
Kramers partners are orthogonal, $\big< \Phi_k \big| \widebar\Phi_k \big> = 0$,
because the only complex number that is equal to its negative value is zero.
QED.

%
%
\section{}
\label{app:diag-GT}
\noindent{\bf Theorem}: The real linear combination of spin matrices have the
following eigenvalues and eigenvectors
\begin{gather} \label{eq:eigen:S}
  \left[ \mb{S}_u, \mb{S}_v \right] = i \epsilon_{uvw} \mb{S}_w,
  \quad
  \left( b_u \mb{S}_u \right) \mb{C} = \mb{C} \mb{e},
  \quad
  \vec b \in \mathbb{R}^3
  \\
  \label{eq:eigen:S:1}
  \Rightarrow
  \quad
  e_k = M_k \big| \,\vec b\, \big|,
  \quad
  M_k \in \{S, S-1, \dots, -S\},
  \quad
  k = 1,\dots,2S+1,
  \\
  \label{eq:eigen:S:2}
  \mb{C} = e^{-i\theta \vec{\mb{S}} \cdot \vec n},
  \quad
  \theta = \mr{cos}^{-1}\left(\frac{b_z}{\big| \,\vec b\, \big|}\right),
  \quad
  \vec n = \left(\big| \,\vec b\, \big|^2 - b_z^2 \right)^{-\frac{1}{2}}
           \left( \vec b \times \vec z\right),
  \quad
  \big| \,\vec{b}\,\big| \ne 0,
\end{gather}
where $\epsilon_{uvw}$ is the Levi--Civita symbol, $S$ is a (half-)integer
number, $\vec z$ represents the unit vector in the z direction, $\mb{C}$ are
eigenvectors in columns, and $\mb{e}$ is a diagonal matrix with the diagonal
elements $e_k$. Note that for $\big| \,\vec{b}\, \big|=0$,
eq~\eqref{eq:eigen:S} has a trivial solution.

\noindent{\bf Proof}:
First we transform the eigenvalue equation~\eqref{eq:eigen:S} by employing the
unitary transformation from SU$(m)$ (with $m=2S+1$)
\begin{gather}
  e^{i\theta \vec{\mb{S}} \cdot \vec n}
  \left( b_u \mb{S}_u \right)
  e^{-i\theta \vec{\mb{S}} \cdot \vec n}
  e^{i\theta \vec{\mb{S}} \cdot \vec n}
  \mb{C}
  = 
  e^{i\theta \vec{\mb{S}} \cdot \vec n}
  \mb{C}\mb{e}.
\end{gather}
Note that only for $m=2$ do these unitary transformations represent all
elements of SU$(m)$, although for higher $m$ they form the $m$-dimensional
irreducible representation of SU$(2)$ [which is a subgroup of SU$(m)$]. Then,
because the spin matrices satisfy the commutation relation in
eq~\eqref{eq:eigen:S}, we can utilize the result of Appendix~\ref{app:rot}
[eq~\eqref{eq:eXeRX} or~\eqref{SU:SO}]
\begin{gather}
  \left[ b_u \left( e^{-i\theta \vec{\mb{R}} \cdot \vec n} \right)_{uv} \mb{S}_v \right]
  e^{i\theta \vec{\mb{S}} \cdot \vec n}
  \mb{C}
  =
  e^{i\theta \vec{\mb{S}} \cdot \vec n}
  \mb{C}\mb{e}
  \\
  \label{eq:tilde-ev}
  \left( \tilde{b}_u \mb{S}_u \right)
  \mb{\tilde{C}}
  =
  \mb{\tilde{C}}\mb{e},
  \\
  \label{eq:tilde-ev-1}
  \mb{\tilde{C}}
  =
  e^{i\theta \vec{\mb{S}} \cdot \vec n}
  \mb{C},
  \\
  \tilde{b}_u = \left( e^{-i\theta \vec{\mb{R}} \cdot \vec n} \right)_{vu} b_v
              = \left( e^{i\theta \vec{\mb{R}} \cdot \vec n} \right)_{uv} b_v.
\end{gather}
Because the matrices $\mr{exp}(i\theta \vec{\mb{R}} \cdot \vec n)$ represent all
proper rotations---all elements of SO$(3)$---the length of $\vec{b}$ and
$\vec{\tilde{b}}$ is the same, $\big| \,\vec{\tilde{b}}\, \big| = \big| \,\vec
b\, \big|$. Moreover, it is always possible to find a proper rotation of
$\vec{b}$ such that $\vec{\tilde{b}}$ will face in the z direction
\begin{equation}
  \label{eq:tildeb}
  \vec{\tilde{b}}
  \overset{!}{=}
  \left(\begin{array}[c]{c}
  0  \\
  0  \\
  \big| \,\vec b\, \big|  \\
  \end{array}\right),
\end{equation}
providing the length of $\vec{b}$ is nonzero. To find the proper
rotation---defined by $\vec n$ and $\theta$---which leads to
eq~\eqref{eq:tildeb}, one must realize that $\vec n$ is a unit vector
orthogonal to both vectors $\vec{b}$ and $\vec{\tilde{b}}$, and that $\theta$
is an angle between these two vectors. Thus by using standard vector algebra we
get two equations
\begin{gather}
  \vec b \cdot \vec{\tilde{b}} = \big| \,\vec b\, \big|^2 \mr{cos}\theta,
  \\
  \vec n = \left( \big| \,\vec b\, \big|^2 \mr{sin}\theta \right)^{-1}
           \left( \vec b \times \vec{\tilde{b}} \right),
\end{gather}
whose solution leads to $\vec n$ and $\theta$ in eq~\eqref{eq:eigen:S:2}.
Finally, if eq~\eqref{eq:tildeb} is satisfied, then eq~\eqref{eq:tilde-ev}
becomes
\begin{gather}
  \left( \big| \,\vec b\, \big| \mb{S}_z \right)
  \mb{\tilde{C}}
  =
  \mb{\tilde{C}}\mb{e},
\end{gather}
and because the spin matrix $\mb{S}_z$ is diagonal with the diagonal elements
\begin{gather}
  S_{kk} \in \{S, S-1, \dots, -S\},
  \quad k = 1,\dots,2S+1,
\end{gather}
the eigenvectors $\mb{\tilde{C}}$ have a simple form, $\tilde{C}_{lk} =
\delta_{lk}$. One thus obtains the eigenvalues $e_k$ from
eq~\eqref{eq:eigen:S:1}, and to get the eigenvectors $\mb{C}$ one uses
eq~\eqref{eq:tilde-ev-1}
\begin{equation}
  C_{lk}
  =
  \left( e^{-i\theta \vec{\mb{S}} \cdot \vec n} \right)_{lp}
  \tilde{C}_{pk}
  =
  \left( e^{-i\theta \vec{\mb{S}} \cdot \vec n} \right)_{lk}.
\end{equation}
QED.

%
%
\section{}
\label{app:proper}

Because the spin operators $\mb{S}_u$ that act in $m$-dimensional complex
space $\mathbb{C}^m$ satisfy the commutation relation in eq~\eqref{XY-YX}
\begin{equation}
  \left[ \mb{S}_u, \mb{S}_v \right] = i \epsilon_{uvw} \mb{S}_w,
\end{equation}
the expression in eq~\eqref{eq:eXeRX} becomes
\begin{align}
  \label{SU:SO}
  e^{i\theta \vec{\mb{S}} \cdot \vec n} \, \mb{S}_u \, e^{-i\theta \vec{\mb{S}} \cdot \vec n}
  =
  \left( e^{-i\theta \vec{\mb{R}} \cdot \vec n} \right)_{uv} \mb{S}_v.
\end{align}
To simplify the notation in this appendix we write eq~\eqref{SU:SO} in the form
\begin{align}
  \label{SU:SO:2}
  \mb{U}^+ \,\mb{S}_u \brr{\mb{U}^+}^\dagger
  =
  \brr{O^+}_{uv} \mb{S}_v,
\end{align}
with the matrices $\mb{O}^+$ and $\mb{U}^+$ being
\begin{align}
  \label{eq:ORn}
  \mb{O}^+ &= e^{-i\theta \vec{\mb{R}}\cdot \vec n},
  \\
  \label{eq:USn}
  \mb{U}^+ &= e^{i\theta \vec{\mb{S}}\cdot \vec n}.
\end{align}
The operator $\mb{O}^+$ is a proper rotation in a real three-dimensional vector
space---{\it i.e.} it is a member of the special orthogonal group
SO$(3)$---while the operator $\mb{U}^+$ is a unitary transformation in a
complex $m$-dimensional vector space---{\it i.e.} it is a member of the special
unitary group SU$(m)$.  Thus both matrices, $\mb{O}^+$ and $\mb{U}^+$, have a
determinant equal to $1$.  Importantly, the expression $\mr{exp}(-i\theta
\vec{\mb{R}}\cdot \vec n)$ represents all group elements of $\mr{SO}(3)$,
while, on the other hand, the expression $\mr{exp}(i\theta \vec{\mb{S}}\cdot
\vec n)$ represents all group elements of SU$(m)$ only in the two-dimensional
case, $m=2$. Note, however, that $\mb{U}^+$ represents all elements of the
$m$-dimensional irreducible representation of SU$(2)$---the subgroup of
SU$(m)$. In the physics literature, this represenation is labeled using $S$
rather than $m$, with $m=2S+1$.

In the rest of this appendix we will discuss the case $m=2$, for which
eq~\eqref{SU:SO} holds for all elements of the groups SU$(2)$ and SO$(3)$.  The
role of the unitary transformation in eq~\eqref{SU:SO}---in the context of this
work---stems from the choice of the orthonormal basis in which the qm operators
are represented [see the discussion in the main text, {\it e.g.}
eq~\eqref{eq:HbarVHV}]. Therefore, in the general case, one must deal with the
expression $\mb{U} \,\mb{S}_u \brr{\mb{U}}^\dagger$ with $\mb{U} \in
\mr{U}(2)$, rather than $\mb{U}^+ \,\mb{S}_u \brr{\mb{U}^+}^\dagger$ with
$\mb{U}^+ \in \mr{SU}(2)$. Note that $\mr{SU}(2)$ is a subgroup of
$\mr{U}(2)$.  However, as we show in the following theorem, the additional
flexibility in $\mb{U}$, as compared to $\mb{U}^+$, is redundant in the
context of eq~\eqref{SU:SO}.

\noindent{\bf Theorem}: For every $\mb{U} \in \mr{U}(2)$ there exist $\mb{U}^+
\in \mr{SU}(2)$ and $\mb{O}^+ \in \mr{SO}(3)$ such that
\begin{gather}
  \label{SU:SO:3}
  \mb{U} \,\mb{S}_u \brr{\mb{U}}^\dagger
  =
  \brr{O^+}_{uv} \mb{S}_v,
  \\
  \label{SU:SO:4}
  \mb{U}^+ \,\mb{S}_u \brr{\mb{U}^+}^\dagger
  =
  \brr{O^+}_{uv} \mb{S}_v,
  \\
  \label{SU:SO:5}
  \mb{U} = e^{i\frac{\alpha}{2}}\mb{U}^+,
  \quad
  \mr{det}\,\mb{U} = e^{i\alpha},
  \quad
  \alpha \in \mathbb{R},
\end{gather}
with $\mb{O}^+$ and $\mb{U}^+$ as defined in eqs~\eqref{eq:ORn}
and~\eqref{eq:USn}.

\noindent{\bf Proof}: Because $\mb{U}^\dagger\mb{U}=\mb{1}$, the determinant of
$\mb{U}$ is a phase factor
\begin{gather}
  \label{phase:U:1}
  \mr{det}(\mb{U}^\dagger\mb{U}) = \mr{det}(\mb{1}),
  \\
  \label{phase:U:2}
  (\mr{det}\,\mb{U})^\ast \,\,\mr{det}\,\mb{U} = 1,
  \\
  \label{phase:U:3}
  \brp{\mr{det}\,\mb{U}}^2 = 1,
  \\
  \label{phase:U:4}
  \Rightarrow
  \quad
  \mr{det}\,\mb{U} = e^{i\alpha}, \quad \alpha \in \mathbb{R}.
\end{gather}
Then for every $\mb{U}$ there exists $\mb{U}^+$ such that
\begin{gather}
  \label{decompose:U}
  \mb{U} = \mb{U}^+ \mb{I}^-,
  \\
  \left(I^-\right)_{kl} = \delta_{kl} \,e^{i\frac{\alpha}{2}}.
\end{gather}
One may readily verify that the determinant of such a $\mb{U}^+$ is one.
Because $\mb{I}^-$ is the identity matrix multiplied by a phase factor, one can
easily modify eq~\eqref{SU:SO:3} to give
\begin{gather}
  \brr{\mb{U}^+\mb{I}^-} \mb{S}_u \brr{\mb{U}^+\mb{I}^-}^\dagger
  =
  \left( O^+ \right)_{uv} \mb{S}_v,
  \\
  \mb{U}^+ e^{i\frac{\alpha}{2}} \mb{S}_u e^{-i\frac{\alpha}{2}} \brr{\mb{U}^+}^\dagger
  =
  \left( O^+ \right)_{uv} \mb{S}_v,
\end{gather}
thus obtaining expression~\eqref{SU:SO:4}. QED.

As a direct consequence of the above theorem, there does not exist a unitary
transformation $\mb{U} \in \mr{U}(2)$ that gives an improper rotation,
$\mb{O}^-$ (a member of O$(3)$ with the determinant equal to minus one), {\it
i.e.}
\begin{align}
  \label{improper:R}
  \mb{U} \mb{S}_u \brr{\mb{U}}^\dagger
  \ne
  \left( O^- \right)_{uv} \mb{S}_v.
\end{align}
The consequence of this statement is that it is impossible to measure the sign
of the individual g-tensor eigenvalues, and only the sign of the g-tensor
determinant is an observable quantity (see the main text for a more detailed
discussion of this fact).

Note that the theorem presented in this appendix is valid only for $m=2$
because $\mb{U}^+$ represents all elements of SU$(2)$ and thus every $\mb{U}
\in \mr{U}(2)$ can be decomposed as presented in eq~\eqref{decompose:U}. For
$m>2$ one may still decompose every $\mb{U} \in \mr{U}(m)$ in a similar way,
except for the fact that the resulting $\mb{U}^+$ , with $\mr{det}\,\mb{U}^+ =
1$, does not necessarily satisfy the relation in eq~\eqref{eq:USn}, and thus
not every such $\mb{U}^+$ satisfies eq~\eqref{SU:SO:4}.

%
%
\section{}
\label{app:rot}

Let us consider three operators $X_u$ on a vector space V which satisfy the
following commutation relation
\begin{gather}
  \left[ \tilde{X}_u, \tilde{X}_v \right] = \epsilon_{uvw} \tilde{X}_w,
\end{gather}
where $\epsilon_{uvw}$ is the Levi--Civita symbol. The three operators
$\tilde{X}_u$ are basis elements of the Lie algebra $\mathfrak{so}(3)$
represented on a vector space V.  To remain consistent with the definition of
momentum operators in quantum theory, we define a new set of operators
$X_u=i\tilde{X}_u$ which then satisfy the commutation relation
\begin{equation} \label{XY-YX}
  \left[ X_u, X_v \right] = i \epsilon_{uvw} X_w.
\end{equation}

\noindent{\bf Theorem}: For the operators $X_u$ that satisfy the commutation
relation~\eqref{XY-YX} the following relations hold
\begin{align} \label{rotX}
  e^{i\theta \vec X \cdot \vec n} \, \vec X \, e^{-i\theta \vec X \cdot \vec n}
  &=
    \vec X \mr{cos}\,\theta + \left( \vec n \times \vec X \right) \mr{sin}\,\theta
  + \left( 1 - \mr{cos}\,\theta \right) \vec n \left( \vec X \cdot \vec n \right),
  \\ 
  \label{eq:eXeRX}
  e^{i\theta \vec X \cdot \vec n} \, X_u \, e^{-i\theta \vec X \cdot \vec n}
  &=
  \left( e^{-i\theta \vec R \cdot \vec n} \right)_{uv} X_v,
\end{align}
where $u, v, w$ are Cartesian indices, $\theta \in \mathbb{R}$, $\vec n$ is a
unit vector in $\mathbb{R}^3$, and $-i\vec R$ are basis elements of the Lie
algebra $\mathfrak{so}(3)$ represented on vector space $\mathbb{R}^3$
\begin{gather}
  (R_u)_{vw} = -i\epsilon_{uvw},
  \label{R:def1}
  \\
  R_1 =
  \left(\begin{array}[c]{ccc}
  0 & 0 & 0 \\
  0 & 0 &-i \\
  0 & i & 0 \\
  \end{array}\right),
  \quad
  R_2 =
  \left(\begin{array}[c]{ccc}
  0 & 0 & i \\
  0 & 0 & 0 \\
 -i & 0 & 0 \\
  \end{array}\right),
  \quad
  R_3 =
  \left(\begin{array}[c]{ccc}
  0 &-i & 0 \\
  i & 0 & 0 \\
  0 & 0 & 0 \\
  \end{array}\right).
  \label{R:def2}
\end{gather}
\noindent{\bf Proof}: To prove expression \eqref{rotX} we employ the
Baker–Campbell–Hausdorff formula
\begin{align}
  e^{i\theta \vec X \cdot \vec n} \vec X e^{-i\theta \vec X \cdot \vec n}
  &= 
  \vec X +
  \left( i\theta \right) \left[ \vec X \cdot \vec n, \vec X \right] +
  \frac{1}{2!} \left( i\theta \right)^2
  \left[ \vec X \cdot \vec n, \left[ \vec X \cdot \vec n, \vec X \right] \right]
  \nonumber
  \\
  &+ 
  \frac{1}{3!} \left( i\theta \right)^3
  \left[
      \vec X \cdot \vec n,
      \left[ \vec X \cdot \vec n, \left[ \vec X \cdot \vec n, \vec X \right] \right]
  \right]
  + \dots.
  \label{transX}
\end{align}
Taking advantage of the commutation relation \eqref{XY-YX} we can write
\begin{gather}
  \left[ \vec X \cdot \vec n, \vec X \right] = - i \left( \vec n \times \vec X \right),
  \label{commut1}
  \\
  \left[ \vec X \cdot \vec n, \left[ \vec X \cdot \vec n, \vec X \right] \right]
  =
  \vec X - \left( \vec X \cdot \vec n \right) \vec n,
  \\
  \left[
      \vec X \cdot \vec n,
      \left[ \vec X \cdot \vec n, \left[ \vec X \cdot \vec n, \vec X \right] \right]
  \right]
  =
  - i \left( \vec n \times \vec X \right).
  \label{commut3}
\end{gather}
Because the commutator expressions, eqs \eqref{commut1} and \eqref{commut3},
are equal, the equation \eqref{transX} simplifies significantly
\begin{align}
  e^{i\theta \vec X \cdot \vec n} \vec X e^{-i\theta \vec X \cdot \vec n}
  &=
  \vec X
  -i
  \sum_{n=0}^{\infty}
      \frac{\left( i\theta \right)^{2n+1}}{\left(2n+1\right)!}
      \left( \vec n \times \vec X \right)
  +
  \sum_{n=1}^{\infty}
      \frac{\left( i\theta \right)^{2n}}{\left(2n\right)!}
      \left( \vec X - \left( \vec X \cdot \vec n \right) \vec n \right)
  \\
  &=
  \sum_{n=0}^{\infty}
      \frac{\left(i\theta\right)^{2n}}{\left(2n\right)!}
      \,\vec X
  +
  \sum_{n=0}^{\infty}
      \frac{\left(-1\right)^{n} \theta^{2n+1}}{\left(2n+1\right)!}
      \left( \vec n \times \vec X \right)
  -
  \sum_{n=1}^{\infty}
      \frac{\left( i\theta \right)^{2n}}{\left(2n\right)!}
      \left( \vec X \cdot \vec n \right) \vec n.
\end{align}
Recognizing the Taylor series of trigonometric functions $\mr{sin}\,\theta$
and $\mr{cos}\,\theta$ one arrives at the expression \eqref{rotX}.

The right-hand-side of the eq~\eqref{rotX} is Rodrigues' well-known rotation
formula. Rodrigues' rotation formula provides an algorithm to compute the
exponential map from the Lie algebra $\mathfrak{so}(3)$ to the Lie group of all
proper rotations, SO$(3)$. If we consider the representation of the Lie algebra
and Lie group on the three-dimensional real vector space $\mathbb{R}^3$, the
Rodrigues' formula then rotates an element of this vector space, $\vec v \in
\mathbb{R}^3$, by given axis $\vec n$ and angle $\theta$
\begin{align} \label{rot:R3}
  e^{-i\theta \vec R \cdot \vec n} \,\vec v = 
    \vec v \, \mr{cos}\,\theta + \left( \vec n \times \vec v \right) \mr{sin}\,\theta
  + \left( 1 - \mr{cos}\,\theta \right) \vec n \left( \vec v \cdot \vec n \right).
\end{align}
Expression~\eqref{rotX} then connects the unitary transformation of the
operators $X_u$ with a proper rotation in $\mathbb{R}^3$ as
\begin{align}
  \label{eq:eXeRX-1}
  e^{i\theta \vec X \cdot \vec n} \vec X e^{-i\theta \vec X \cdot \vec n}
  =
  e^{-i\theta \vec R \cdot \vec n} \,\vec X,
\end{align}
which equals to eq~\eqref{eq:eXeRX}. QED.

In eq~\eqref{eq:eXeRX-1} the operator $O(\theta, \vec n) = exp(-i\theta \vec R
\cdot \vec n)$ represents all proper rotations in $\mathbb{R}^3$, {\it i.e.}
all rotations with $\mr{det}(O) = 1$.  Note that operators $X_u$ can have
different forms depending on what vector space is chosen for the representation
of the lie algebra $\mathfrak{so}(3)$. This includes, but is not limited to,
the operators
\begin{gather}
  (R_u)_{vw} = -i\epsilon_{uvw},
  \\
  \vec L = -i \left( \vec r \times \vec\nabla \right),
  \\
  \vec S = \frac{1}{2} \vec\sigma,
  \\
  \vec J = \vec L + \vec S,
\end{gather}
with $\vec\sigma$ representing a vector composed of Pauli matrices.

%
%
\section{}
\label{app:eff:spin}
In this appendix we discuss the definition of the fictitious spin space and its
consequences. We will demonstrate the ideas using the case of the doublet
system, but the theory can easily be extended to any arbitrary multiplicity.
The basic idea of the fictitious spin space is to represent wavefunctions in
the complex vector space in a way that preserves their properties.
Many-electron wavefunctions belong to the Fock subspace, which is equipped with
the inner product $\bapa{\Phi_1}{\Phi_2} \rightarrow \mathbb{C}$, see also
Appendix~\ref{app:K}.  These wavefunctions are then associated with vectors in
the complex vector space $\mathbb{C}^m$---also known as spin vectors---equipped
with the dot product $\vec{v} \cdot \vec{w} = \mb{v}^\dagger \mb{w} \rightarrow
\mathbb{C}$. To associate a wavefunction with a vector from the fictitious spin
space, one first selects a reference basis set $\{\Phi_n\}_{n=1}^m$ and expands
the wavefunction in it, $\Psi = C_n \Phi_n$.  The wavefunction is then
represented by the expansion coefficients used to construct the vector in
$\mathbb{C}^m$, $v_n \coloneqq C_n$.  In the doublet case, $m=2$, the reference
basis set itself--- the Kramers pair $\{\Phi,\widebar\Phi\}$---is thus
associated with the two-dimensional spin vectors in the following manner
\begin{align}
  \label{eq:psi-v-equiv}
  \Phi
  \rightarrow
  \mb{v}
  =
  \left(\begin{array}[c]{c}
  1 \\
  0
  \end{array}\right),
  \qquad
  \widebar\Phi
  \rightarrow
  \widebar{\mb{v}}
  =
  \left(\begin{array}[c]{c}
  0 \\
  1
  \end{array}\right).
\end{align}
As discussed in section~\ref{sec:introduction}, the purpose of the EPR
effective spin Hamiltonian is to describe projection of the qm Hamiltonian onto
a space defined by a set of wavefunctions $\{\Phi_n\}_{n=1}^m$, {\it i.e.}
$\mb{H}^\mr{eff} \overset{!}{=} \mb{H}^\mr{qm}$. For the doublet case this
equivalence is presented in eq~\eqref{gt:def1}. The mapping in
eq~\eqref{eq:psi-v-equiv} then leads to the same expectation value---either of
the wavefunction with the Zeeman Hamiltonian or of the spin vector with the
effective spin Hamiltonian---regardless of the type of space it is calculated
in. Thus for example for the basis wavefunction $\Phi$ it holds that
\begin{align}
  \label{eq:psi-v-H}
  \bappa{\Phi}{H^Z}{\Phi}
  \quad
  \rightarrow
  \quad
  \Big( \:1 \quad 0 \:\Big)^\ast
  &\left(
    \begin{array}{cc}
        &          \\
      \multicolumn{2}{c}{\smash{\raisebox{.5\normalbaselineskip}{$\frac{1}{2c} B_u g_{uv} \mb{S}_v$}}}
                   \\
    \end{array}
  \right)
  \left(\begin{array}[c]{c}
  1 \\
  0
  \end{array}\right)
  \\
  =
  \Big( \:1 \quad 0 \:\Big)^\ast
  &\left(\begin{array}[c]{cr}
  \big<         \Phi \big| H^Z \big| \Phi \big> & \big< \Phi \big| H^Z \big| \widebar\Phi \big> \\
  \big< \widebar\Phi \big| H^Z \big| \Phi \big> &-\big< \Phi \big| H^Z \big|         \Phi \big>
  \end{array}\right)
  \left(\begin{array}[c]{c}
  1 \\
  0
  \end{array}\right),
\end{align}
with $\vec{\mb{S}} = \tfrac{1}{2}\bs{\sigma}$.  Eq~\eqref{eq:psi-v-H}
represents an expectation value for one basis function; the expression for all
combinations of basis functions can be written as follows
\begin{align}
  \label{eq:psi-v-H-gen}
  \bappa{\Phi_m}{H^Z}{\Phi_n}
  \quad
  \rightarrow
  \quad
  \mb{v}_m^\dagger
   \left( \frac{1}{2c} B_u g_{uv} \mb{S}_v \right)
  \mb{v}_n,
\end{align}
with $\{\Phi_1,\Phi_2\} = \{\Phi,\widebar\Phi\}$ and $\{\mb{v}_1,\mb{v}_2\} =
\{\mb{v},\widebar{\mb{v}}\}$. Choosing a different orthonormal Kramers pair as
a basis set for representation of the Zeeman Hamiltonian corresponds to the
unitary transformation $\widetilde\Phi_m = U_{nm} \Phi_n$.  In the main text we
have discussed the usefulness of the unitary transformations that belong to the
$m$-dimensional irreducible representations of the SU$(2)$ group [in the case
of $m=2$ this is the natural representation of SU$(2)$ in $\mathbb{C}^2$]
\begin{gather}
  \label{eq:U-phi-trans-1}
  \widetilde{\Phi}_m
  =
  \left( e^{-i \theta \vec{\mb{S}} \cdot \vec n } \right)_{nm} \Phi_n.
\end{gather}
One of the advantages of such transformations is that if the set
$\{\Phi_n\}_{n=1}^m$ satifies relations in eq~\eqref{eq:TRrelations}, then the
set $\{\widetilde{\Phi}_n\}_{n=1}^m$ fulfills these relations as well [see also
the corresponding discussion under eq~\eqref{eq:TRrelations}]. Therefore, for
$m=2$, the basis set on the left-hand-side of eq~\eqref{eq:U-phi-trans-1} forms
a Kramers pair as well. According to eq~\eqref{eq:U-phi-trans-1} the spin
vectors $\widetilde{\mb{v}}_m$ associated with the basis functions
$\widetilde{\Phi}_m$ can be written as $(\widetilde{v}_m)_n = [\mr{exp}(-i
\theta \vec{\mb{S}} \cdot \vec n)]_{nm}$, or equivalently
\begin{align}
  \label{eq:psi-v-equiv1}
  \widetilde{\Phi}
  \rightarrow
  \widetilde{\mb{v}} =
  \left(
    \begin{array}{cc}
        &          \\
      \multicolumn{2}{c}{\smash{\raisebox{.5\normalbaselineskip}{$e^{-i \theta \vec {\mb{S}} \cdot \vec n }$}}}
                   \\
    \end{array}
  \right)
  \left(\begin{array}[c]{c}
  1 \\
  0
  \end{array}\right),
  \qquad
  \widetilde{\widebar\Phi}
  \rightarrow
  \widebar{\widetilde{\mb{v}}} =
  \left(
    \begin{array}{cc}
        &          \\
      \multicolumn{2}{c}{\smash{\raisebox{.5\normalbaselineskip}{$e^{-i \theta \vec {\mb{S}} \cdot \vec n }$}}}
                   \\
    \end{array}
  \right)
  \left(\begin{array}[c]{c}
  0 \\
  1
  \end{array}\right),
\end{align}
where $\{\widetilde{\Phi}_1, \widetilde{\Phi}_2\} = \{\widetilde{\Phi},
\widebar{\widetilde{\Phi}}\}$ and
$\{\widetilde{\mb{v}}_1,\widetilde{\mb{v}}_2\} =
\{\widetilde{\mb{v}},\widebar{\widetilde{\mb{v}}}\}$.  Employing
eqs~\eqref{eq:U-phi-trans-1} and~\eqref{SU:SO} one may calculate the
expectation values of the set $\{\widetilde{\Phi},
\widetilde{\widebar{\Phi}}\}$ with the Zeeman operator in both the real and
fictitious spin spaces as
\begin{gather}
  \label{eq:psi-v-H-new}
  \bappa{\widetilde{\Phi}_m}{H^Z}{\widetilde{\Phi}_n}
  =
  \brr{e^{i \theta \vec {\mb{S}} \cdot \vec n }}_{\!mk}
  \bappa{\Phi_k}{H^Z}{\Phi_l}
  \brr{e^{-i \theta \vec {\mb{S}} \cdot \vec n }}_{\!ln}
  \\
  \rightarrow
  \\
  \label{eq:tilde-eff-1}
  \widetilde{\mb{v}}^\dagger_m
  \brr{ \frac{1}{2c} B_u g_{uv} \mb{S}_v }
  \widetilde{\mb{v}}_n
  \\
  \label{eq:tilde-eff-2}
  =
  \brr{e^{i \theta \vec {\mb{S}} \cdot \vec n}}_{\!mk}
  \brr{ \frac{1}{2c} B_u g_{uv} \mb{S}_v }_{\!kl}
  \brr{e^{-i \theta \vec {\mb{S}} \cdot \vec n}}_{\!ln}
  \\
  \label{eq:tilde-eff-3}
  =
  \left(
    \frac{1}{2c} B_u g_{uv}\,
    e^{i \theta \vec {\mb{S}} \cdot \vec n }
    \mb{S}_v
    e^{-i \theta \vec {\mb{S}} \cdot \vec n }
  \right)_{\!mn}
  \\
  \label{eq:tilde-eff-4}
  =
  \mb{v}_m^\dagger
  \left(
    \frac{1}{2c} B_u g_{uv}\,
    \widetilde{\mb{S}}_v
  \right)
  \mb{v}_n
  \\
  \label{eq:tilde-eff-5}
  =
  \mb{v}_m^\dagger
  \left[
    \frac{1}{2c} B_u g_{uv}
    \left( e^{-i \theta \vec{\mb{R}} \cdot \vec n } \right)_{uw} \mb{S}_w
  \right]
  \mb{v}_n
  \\
  \label{eq:tilde-eff-6}
  =
  \mb{v}_m^\dagger \left( \frac{1}{2c} B_u \widetilde{g}_{uv} \mb{S}_v \right) \mb{v}_n.
\end{gather}
Finally, one may calculate the orientation of the fictitious spin of
wavefunctions $\Phi$ and $\widetilde\Phi$ as follows
\begin{align}
  \label{eq:rot-spin-1}
  \vec{s} &= \mb{v}^\dagger \vec{\mb{S}} \,\mb{v},
  \\
  \label{eq:rot-spin-2}
  \vec{\tilde{s}}
  &= \widetilde{\mb{v}}^\dagger \vec{\mb{S}} \,\widetilde{\mb{v}}
  \\
  \label{eq:rot-spin-3}
  &=
  \widetilde{\mb{v}}^\dagger
  e^{i \theta \vec {\mb{S}} \cdot \vec n }
  \,\vec{\mb{S}}
  \,e^{-i \theta \vec {\mb{S}} \cdot \vec n }
  \,\widetilde{\mb{v}},
  \\
  \label{eq:rot-spin-4}
  &=
  \widetilde{\mb{v}}^\dagger
  e^{-i \theta \vec{\mb{R}} \cdot \vec n }
  \,\vec{\mb{S}}
  \,\widetilde{\mb{v}},
  \\
  \label{eq:rot-spin-5}
  &=
  e^{-i \theta \vec{\mb{R}} \cdot \vec n }
  \vec{s}.
\end{align}
One may deduce a few facts from the discussion in this appendix:
\begin{enumerate}
\item
The first important observation is that the orientation of the fictitious spin
of the reference basis wavefunctions in the fictitious spin space, see
eqs~\eqref{eq:psi-v-equiv} and~\eqref{eq:rot-spin-1}, is always along the
z-axis.
\item
From point 1 one infers that by choosing the reference basis set one also
chooses the axis system in the fictitious spin space.
\item
Again from point 1, one may see that the orientation of the real and fictitious
spin are not related unless additional assumptions are imposed.
\item
From eqs~\eqref{eq:psi-v-H-gen},~\eqref{eq:tilde-eff-1},
and~\eqref{eq:rot-spin-5} one concludes that choosing a different basis set of
orthonormal wavefunctions is equivalent to rotation of the fictitious spin of
those wavefunctions.
\item
Eqs~\eqref{eq:tilde-eff-1} and~\eqref{eq:tilde-eff-4} correspond to active
and passive rotation, respectively, where the active transformation rotatates
spin vectors in a fixed coordinate axis system and the passive transformation
rotates the axis system---thus changing the spin operators---while the spin
vectors remain fixed.
\item
Finally, eqs~\eqref{eq:tilde-eff-5} and~\eqref{eq:tilde-eff-6} correspond to
rotation of the g-tensor while leaving the spin vectors and axis system
unchanged.
\end{enumerate}

%
%
\section{}
\label{app:U}

\noindent{\bf Theorem}:
Let us consider an $m$-dimensional vector space $V$ over a field $F$ of either
real $\mathbb{R}$ or complex $\mathbb{C}$ numbers, with a defined dot product
of two vectors $\vec{v}, \vec{w} \in V$ such that $\vec{v} \cdot \vec{w}
\rightarrow F$.  The generalization of the dot product to abstract vector
spaces is called the inner product, see also
eqs~\eqref{eq:ip-prop-0}--\eqref{eq:ip-prop-3}.  Then, for every set of the
orthonormal vectors $\{\vec{w}_v\}_{v=1}^m$ from the vector space $V$, it holds
that
\begin{gather}
  \label{eq:wdotwuv}
  \vec{w}_u \cdot \vec{w}_v = \delta_{uv}
  \quad
  \Rightarrow
  \quad
  \mb{W}^\mathsf{T} \mb{W} = \mb{W} \mb{W}^\mathsf{T} = \mb{1},
\end{gather}
where the matrix $\mb{W}$ is composed of the coordinates of the vectors
$\vec{w}_v$ in columns
\begin{gather}
  (w_v)_u \coloneqq W_{uv},
\end{gather}
with $(w_v)_u$ being defined using the orthonormal basis set
$\{\vec{n}_u\}_{u=1}^m$ such that
\begin{gather}
  \label{eq:wdotn}
  \vec{w}_v \cdot \vec{n}_u = (w_v)_u,
  \\
  \label{eq:ndotn}
  \vec{n}_u \cdot \vec{n}_v = \delta_{uv}.
\end{gather}
Eq~\eqref{eq:wdotn} represents a projection of the vector $\vec{w}_v$ onto the
direction defined by the vector $\vec{n}_u$.  Note that the transpose sign
$\mathsf{T}$ in eq~\eqref{eq:wdotwuv} indicates that we have chosen the field
of real numbers; the theorem and the proof is, however, easily applicable to
the field of complex numbers as well.

\noindent{\bf Proof}: Any vector can be viewed as a linear combination of the
basis vectors with its coordinates being the expansion coefficients
\begin{gather}
  \label{eq:wexpansion}
  \vec{w}_v = (w_v)_k \,\vec{n}_k.
\end{gather}
Eq~\eqref{eq:wdotn} may then be readily verified
\begin{gather}
  \vec{w}_v \cdot \vec{n}_u
  = \brs{ (w_v)_k \,\vec{n}_k } \cdot \vec{n}_u
  = (w_v)_k \brs{ \vec{n}_k \cdot \vec{n}_u }
  = (w_v)_k \,\delta_{ku}
  = (w_v)_u,
\end{gather}
where we have employed the orthonormalization condition for the basis set,
eq~\eqref{eq:ndotn}.  Using the definition of a vector through expansion in
eq~\eqref{eq:wexpansion} and the orthonormalization condition in
eq~\eqref{eq:wdotwuv} one can easily prove the first identity
\begin{gather}
  \vec{w}_u \cdot \vec{w}_v = \delta_{uv},
  \\
  \brs{ (w_u)_k \,\vec{n}_k } \cdot \brs{ (w_v)_l \,\vec{n}_l } = \delta_{uv},
  \\
  (w_u)_k \, \delta_{kl} \, (w_v)_l = \delta_{uv},
  \\
  (w_u)_k \, (w_v)_k = \delta_{uv},
  \\
  W_{ku} \, W_{kv} = \delta_{uv},
  \\
  \mb{W}^\mathsf{T} \mb{W} = \mb{1}.
\end{gather}
To prove the second identity one can simply reverse the role of the two vector
sets, $\{\vec{w}_v\}_{v=1}^m$ and $\{\vec{n}_v\}_{v=1}^m$, where now one
expands the vectors $\vec{n}_v$ in the orthonormal basis set
$\{\vec{w}_v\}_{v=1}^m$ instead
\begin{gather}
  \vec{n}_u \cdot \vec{n}_v = \delta_{uv},
  \\
  \brs{ (n_u)_k \,\vec{w}_k } \cdot \brs{ (n_v)_l \,\vec{w}_l } = \delta_{uv},
  \\
  (n_u)_k \, \delta_{kl} \, (n_v)_l = \delta_{uv},
  \\
  (n_u)_k \, (n_v)_k = \delta_{uv},
  \\
  \label{eq:nnreverse}
  \brs{ \vec{n}_u \cdot \vec{w}_k } \, \brs{ \vec{n}_v \cdot \vec{w}_k } = \delta_{uv},
  \\
  (w_k)_u  \, (w_k)_v = \delta_{uv},
  \\
  W_{uk} \, W_{vk} = \delta_{uv},
  \\
  \mb{W} \mb{W}^\mathsf{T} = \mb{1}.
\end{gather}
To obtain eq~\eqref{eq:nnreverse} we have used eq~\eqref{eq:wdotn} with the
roles of the vectors $\vec{n}_u$ and $\vec{w}_u$, ({\it i.e.} which one is used
as a basis and which is the vector being expanded) reversed. In other words,
$(n_u)_k$ may be calculated as the projection of the vector $\vec{n}_u$ onto
the direction defined by the vector $\vec{w}_k$. QED.

%
%
\section{}
\label{app:kramers-partner}

{\bf Theorem:} If $\Psi$ is an eigenfunction of the time-reversal symmetric
Hamiltonian $H$, then its (non-)Kramers partner, $\widebar{\Psi} \coloneqq
\mc{K}\Psi$, is an eigenfunction of $H$ with the same eigenvalue and the same
norm
\begin{equation}
  \label{eq:KP1}
  H\Psi = E\Psi,
  \quad
  [H,\mc{K}]=0
  \quad
  \Rightarrow
  \quad
  H \widebar\Psi = E \widebar\Psi,
  \quad
  \bapa{\Psi}{\Psi} = \bapa{\widebar\Psi}{\widebar\Psi}.
\end{equation}

\noindent{\bf Proof:} Utilizing the time-reversal symmetry of the Hamiltonian
$H$, the fact that the time-reversal operator is unitary,
eq~\eqref{antilinear1}, a proper definition of the adjoint of the antilinear
operator, eq~\eqref{antilinear3}, and the conjugate symmetry,
eq~\eqref{eq:ip-prop-1}, one can easily show that
\begin{gather}
  H\Psi = E\Psi,
  \\
  \mc{K}H\Psi = \mc{K}E\Psi,
  \\
  H\mc{K}\Psi = E\mc{K}\Psi,
  \\
  \label{eq:psibar4}
  H\widebar{\Psi} = E\widebar{\Psi},
\end{gather}
and
\begin{gather}
  \bapa{\Psi}{\Psi} = \bapa{\Psi}{\mc{K}^\dagger\mc{K}\Psi}
                    = \bapa{\mc{K}\Psi}{\mc{K}\Psi}^\ast
                    = \bapa{\widebar\Psi}{\widebar\Psi}.
\end{gather}
QED. Note that $\Psi$ and $\widebar{\Psi}$ are not necessarily different
wavefunctions, see the case of an even number of electrons discussed in
Appendix~\ref{app:non-magnetic1}.

\end{appendices}

%
%
\providecommand{\latin}[1]{#1}
\makeatletter
\providecommand{\doi}
  {\begingroup\let\do\@makeother\dospecials
  \catcode`\{=1 \catcode`\}=2 \doi@aux}
\providecommand{\doi@aux}[1]{\endgroup\texttt{#1}}
\makeatother
\providecommand*\mcitethebibliography{\thebibliography}
\csname @ifundefined\endcsname{endmcitethebibliography}
  {\let\endmcitethebibliography\endthebibliography}{}

\end{document}